\documentclass[journal]{IEEEtran}
\usepackage{color}
\usepackage{xcolor,colortbl}
\usepackage{graphicx}
\usepackage{tikz}
\usepackage{tikzscale}
\usetikzlibrary{arrows.meta}
\usepackage{pgfplots}
\usepackage{dashrule}
\usepackage{caption}
\usepackage{subcaption}
\usepackage{multicol}
\usepackage{balance}
\usepackage{diagbox}
\usepackage{multirow}
\usepackage{hanging}
\usepackage{ragged2e}
\usepackage{romannum}

\usepackage{listings}

\usepackage{url}
\usepackage{xurl}
\usepackage{hyperref}
\usepackage{cite}

\usepackage{algorithmic}
\usepackage[titlenumbered,ruled]{algorithm2e}

\usepackage{soul}       
\usepackage{textcomp}   
\usepackage{comment}

\usepackage{amsmath,amssymb,amsfonts}

\usepackage{accents}     

\newcounter{substep}

\ifCLASSINFOpdf
\else
\fi
\begin{document}
%
\title{Angle of Arrival Estimation for Gesture Recognition from reflective body-worn tags}
%
%
%
%

\author{Sahar~Golipoor,
        Reza~Ghazalian, Inês~Lobato~Mesquita, and~Stephan~Sigg
\thanks{We acknowledge funding by the European Union in the frame of the Horizon Europe EIC project SUSTAIN (project no. 101071179), and Holden (project no. 101099491). Views and opinions expressed are those of the authors and do not necessarily reflect those of the European Union.}
\thanks{Sahar~Golipoor, Reza Ghazalian, and Stephan~Sigg are with the Department of Information and Communications Engineering, Aalto University, Espoo, 02150 Finland  (e-mail: \{sahar.golipoor, ext-reza.ghazalian, stephan.sigg\}@aalto.fi).}

\thanks{Inês~Lobato~Mesquita is with KTH Royal Institute of Technology, SE-100 44, Stockholm, Sweden, and also the Electrical and Computer Engineering Department at Instituto Superior Técnico (e-mail: inesme@kth.se)}
}

\maketitle
\sethlcolor{blue!50!white}

\begin{abstract}
We investigate hand gesture recognition by leveraging passive reflective tags worn on the body. Considering a large set of gestures, distinct patterns are difficult to be captured by learning algorithms using backscattered received signal strength (RSS) and phase signals. This is because these features often exhibit similarities across signals from different gestures. To address this limitation, we explore the estimation of Angle of Arrival (AoA) as a distinguishing feature, since AoA characteristically varies during body motion.
To ensure reliable estimation in our system, which employs Smart Antenna Switching (SAS), we first validate AoA estimation using the Multiple SIgnal Classification (MUSIC) algorithm while the tags are fixed at specific angles. Building on this, we propose an AoA tracking method based on Kalman smoothing. Our analysis demonstrates that, while RSS and phase alone are insufficient for distinguishing certain gesture data, AoA tracking can effectively differentiate them.
To evaluate the effectiveness of AoA tracking, we implement gesture recognition system benchmarks and show that incorporating AoA features significantly boosts their performance. Improvements of up to 15\% confirm the value of AoA-based enhancement.
\end{abstract}





%
\IEEEpeerreviewmaketitle

\section{Introduction}\label{sec:introduction}
\IEEEPARstart{R}{adio} based human sensing comprises the analysis of electromagnetic signals reflected back from individuals or objects, with the aim of, for instance, localization~\cite{ghazalian2023joint} gesture~\cite{palipana2021pantomime}, vital sign~\cite{jiang2023automatic} or emotion recognition~\cite{alabsi2021emotion}. It is essential part of many applications including remote health monitoring, ambient assisted living, as well as smart home and factory~\cite{xiao2023survey}. Challenges in Radio-based human sensing exist. One primary difficulty lies in the fact that the received signal constitutes a superposition of reflections from both the target subject, such as a human, and various non-target objects present in the environment. This complicates accurate recognition and interpretation of the human-related signal components. Furthermore, since a region of interest is defined for human sensing and everybody in the region could be sensed, one will find radio-based human sensing privacy invasion. 

Concurrently, ultra-high frequency radio frequency identification (UHF RFID) tag, which is conventionally known for identification applications, has demonstrated its potential for sensing thanks to its low cost, contactless modality and capability of making the targets' surfaces intelligent~\cite{landaluce2020review}~\cite{haibi2022systematic}. 
Each RFID tag contains a unique Electronic Product Code (EPC). Therefore, a potential solution to the previously mentioned challenges involves the use of smart clothing embedded with RFID tags.
a) By wearing such garments, it becomes possible to determine the specific body region from which a signal is transmitted during RFID tag-reader interactions. This can enhance gesture recognition, particularly for complex gestures.
b) By wearing such smart cloths, individuals who choose to be monitored can be uniquely identified, while those concerned about privacy will not be detected.

There are different sensing indicators for received signals such as time of arrival (ToA), Time Difference of Arrival (TDoA), received signal strength (RSS), phase of arrival (PoA), and phase difference of arrival (PDoA)~\cite{xu2023principle}.

UHF RFID works in 860-960 MHz frequency range and does not offer high bandwidth. Thus, ToA and TDoA estimation has considerable error and will not help in RFID-based sensing scenarios~\cite{florio2025localization}. Besides, RSS and phase-based approaches are highly influenced via fading and multipath effects~\cite{faseth2011influence} and are not suffeciently robust and reliable in some scenarios. Figs.~\ref{RssGestures} and~\ref{PhaseGestures} , for example, show the extracted RSS and phase received from an RFID tags attached on human hand while doing different gestures. As can be seen they are relatively analogous for two different gestures and might not be solitary useful for recognition purposes. On the other hand, angular measurements are capable to provide unique information for sensing scenarios. Angle of arrival (AoA) in receiver provide knowledge from which direction signals are coming that is advantageous, particularly in target tracking applications, and real-time localization~\cite{pereira2025long}~\cite{fabris2025AoA}. In general, estimation of this indicators would be beneficial specially if they were fused with RSS or phase. Accordingly, we explore AoA tracking within a body-worn RFID system as a means to enhance gesture recognition.
Our contributions are
\begin{itemize}
\item the estimation of AoA of tags in the azimuth direction using the MUSIC algorithm within an RFID system that employs Smart Antenna Switching.
\item an enhanced performance of gesture recognition systems from body-worn reflective tags, using AoA tracking refined with a Kalman filter.
\item a validation of the benefit of AoA for human gesture recognition on various benchmark models.
\end{itemize}

\section{Related Work}\label{related work}
Popular environmental radio sensing modalities are RFID, WiFi and mmWave Radar.
Particularly, the application of sign-language recognition has been prominently investigated for RFID-based recognition systems~\cite{zhang2024sign,xu2023rf,dian2020towards}.
Classical machine learning (e.g. dynamic time warping (DTW), support vector machines (SVM), decision trees, random forest classifiers) have been utilized traditionally~\cite{merenda2022edge,zhang2022real,zou2016grfid}, but also convolutional neural networks (CNN)-based approaches~\cite{wang2018multi} and data augmentation were used with good success~\cite{ma2022rf}.

The channel state information (CSI) in WiFi contains information on the signal strength and the phase of the received signal across multiple frequency carriers.
This information can be utilized for sensing~\cite{regani2022gwrite}, using e.g.  CNN~\cite{yao2023human} or graph-based neural networks~\cite{chen2024wignn}.
A common challenge is the dependency of the recognition on the relative orientation of a subject and the WiFi device, which can be addressed by antenna placement settings~\cite{qin2024direction}.
Attention mechanisms have been successfully employed to improve recognition accuracy through weighing amplitude and phase features differently~\cite{gu2023attention,gu2022wigrunt}.

Radar-based motion recognition has received increased attention recently and predominantly for human gesture recognition~\cite{jin2023interference} exploiting range-angle, range-Doppler, range-time and angle-time maps~\cite{yu2024mmwave, wu2024lightweight}.
In the literature, attention mechanisms~\cite{jin2024rodar}, interpretation of spectogram images via image-based neural networks~\cite{qiao2024simple}, as well as graph-based appraoches~\cite{salami2022tesla} have been proposed to classify radar-data.

A major challenge in the aforementioned works is that the received signal is a superposition of multipath reflections and has to first be separated into the signal of interest versus reflections from environmental objects.
In addition, radio sensing faces significant privacy concerns since, similar to video or audio-based systems, it is not easily possible to constrain the sensing of the system spatially or towards target groups.

We propose attaching reflective tags on-body to mitigate reflections from the environment and other subjects.
Likewise, body-worn systems significantly reduce privacy issues since conscious wearing of sensors comprises an implicit consent mechanism.

RFID has been integrated into objects and tools for studies on human-object interaction or daily activities~\cite{li2015idsense}, customer behavior~\cite{zhou2017design}, capture sentiments associated with specific items~\cite{shangguan2017enabling}, and observe collaborative actions in high-stakes settings like trauma resuscitation~\cite{li2016deep}. In~\cite{bu2018rf}, the authors advanced this concept by attaching RFID arrays to objects and leveraging orthogonal antennas for motion tracking, effectively turning physical items into interactive human-computer interfaces.
Unlike these prior approaches, which typically rely on static RFID infrastructures embedded in the environment or fixed to objects, our work introduces a dynamic system where RFID tags are embedded directly into clothing. This configuration allows the tags to move synchronously with the user’s gestures. Thanks to their minimal weight and flexibility, RFID tags can be attached to garments, making them useful for applications in smart textiles and wearable technology~\cite{salo2024carbon, kruse2024smart}. Moreover, in contrast to environmental setups, wearable RFID systems provide the added benefit of differentiating between movements of distinct body parts, as shown in our previous work~\cite{golipoor2023accurate}.

In~\cite{golipoor2024rfid}, we introduced a wearable RFID-based system capable of recognizing activities related to presence detection by jointly analyzing features derived from both phase and RSS signal. Prior studies have leveraged DTW for fine-grained gesture recognition, particularly through RFID tags embedded in wearable items like gloves~\cite{cheng2019air, xie2017multi}. Additionally, a position-invariant method was presented in~\cite{zhang2023rf}, where sign language recognition was achieved by normalizing key hand motion attributes such as the horizontal rotation angle and radial distance.

Employing RFID tags on the human body for coarse-grained gesture recognition introduces challenges. Frequent changes in tag orientation during movements disrupts stable signal reception by RFID readers and leads to degraded signal quality due to polarization mismatches~\cite{clarke2006radio}. Additionally, signal reflection is affected by anatomical features and tissue composition~\cite{vasisht2018body}. As a result, differences in body shape and tissue can alter reflection patterns, making it difficult to maintain distinguishable signal signatures for the same gesture, thus reducing recognition accuracy. To address these challenges,~\cite{yu2019rfid} proposed a multi-modal strategy for coarse-grained gesture recognition, integrating CNNs and Long Short-Term Memory (LSTM) models to extract and interpret features from RFID data. In our previous work~\cite{golipoor2024environment}, we developed two models, one based on a VGG16 architecture that leverages IQ scatter plots, and another phase-driven approach that applies segmentation via zero-crossing detection combined with a refined derivative-based technique. In contrast to the above studies, we incorporate tags' AoA tracking as new features to facilitate gesture recognition models.

Directional of arrival-based sensing outperforms ToA and TDoA by needing fewer anchors and less synchronization. Diverse techniques and algorithms for AoA finding have been reviewed in~\cite{kulkarni2025comprehensive} in terms of theoretical foundations and computational complexity. Direction of arrival has been studied using WiFi, Radar, UWB, and RFID~\cite{fischer2024systematic}. RFID uses backscatter communication by default, and backscatter-based UWB has also been explored in researches~\cite{mazhar2017precise}. While UWB modules offer accurate timing for tracking and positioning, they require batteries, manual maintenance, and are hard to attach. In contrast, passive UHF RFID tags are wirelessly powered and easily embedded due to their small size and light weight.

Tags designs were conducted to perform AoA and orientation based localization~\cite{abbas2019dielectric, ali2023dual, gil2021frequency}.
Using software defined radio (SDR), AoA was measured for Gen2 tag inventory~\cite{mitsugi2021simultaneous}, and localization~\cite{skyvalakis2022elliptical}~\cite{ma2018indoor}. 
AoA estimation has been conducted by resolving the $\pi$ phase ambiguity~\cite{zhao2024pushing} and array based phase separation technique~\cite{liu2024simultaneous}. In~\cite{wang2022tri}, AoA estimates has been refined with deep learning on 2-dimensional feature image. To determine the direction angle, a retrieval method using coordinate system rotation ~\cite{xie2022AoA}, and mutual coupling interference~\cite{wang2022accurate} were investigated. AoA-based localization has been studied using frequency-scanning leaky-wave antennas~\cite{gil2022direction} and sparse tag arrays~\cite{yang2021rfid}.
Contrary to the mentioned works that employed AoA for positioning, we investigate the use of AoA tracking of wearable tags for gesture recognition.


\subsubsection*{Notation}
 Vectors and matrices are denoted by lowercase and uppercase bold letters, respectively. The $i$-th element of vector $\mathbf{a}$ is represented by $[\mathbf{a}]_{i}$. The transpose operator is denoted by $(\cdot)^\top$, and the conjugate transpose (Hermitian) operator by $(\cdot)^\mathsf{H}$. The $L_2$ norm (Euclidean norm for vectors and Frobenius norm for matrices) is denoted by $\|\cdot\|$. The four-quadrant inverse tangent function is written as $\mathrm{atan2}(y,x)$. The expectation operator is denoted by $\mathbb{E}(\cdot)$, and $\mathbf{I}_2$ denotes the identity matrix of size $2 \times 2$. $\mathrm{Tr}(\mathbf{A})$ denotes the trace of the matrix $\mathbf{A}$, which is the sum of its diagonal elements.

\section{System and Signal Models}\label{sec_sys_sig_model}
In this section, we introduce the RFID system model considered, together with the models of the received signals that will be used in the gesture recognition models.
\subsection{System Model}

\begin{figure}
\centering
\includegraphics[width=0.7\linewidth]{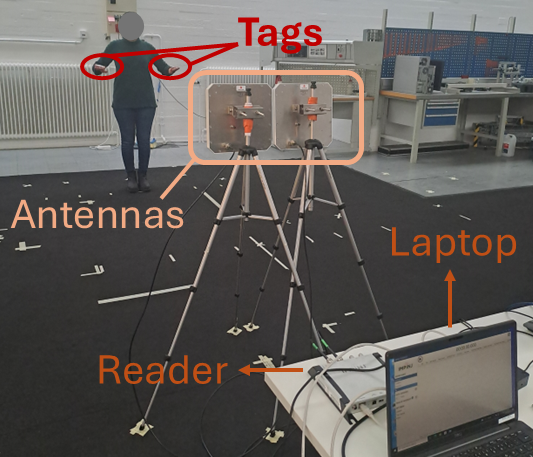}
\caption{Experimental setup for body-worn RFID-based gesture recognition. Two passive RFID tags are attached to the subject, with phase and RSS data collected using an Impinj Speedway R420 reader and two circularly polarized antennas.}
\label{SetUp}
\end{figure}

Consider the RFID system scenario illustrated in Fig~\ref{SetUp}, which consists of a two-element antenna RFID reader located at a known position, $\mathbf{p}{_\text{R}} \in \mathbb{R}^2$. The system also includes one or more passive\footnote{The passive RFID tags are powered by the transmitted signal from the RFID reader.} single-antenna tags located at unknown positions. 
These positions are denoted as $\mathbf{p}{_\text{i}} \in \mathbb{R}^2 $, where $i \in {1, \dots, N_t}$, and $N_t$ denotes the number of tags. The tags must be within the field of view of the RFID antennas to avoid misdetection.\footnote{Tag misdetection also depends on the transmit power, which can be configured through the RFID reader's software.} The RFID reader transmits a carrier wave (CW) using right-hand circular polarization to prevent misdetection caused by polarization mismatch between the RFID reader's antennas and the tags' antennas. Additionally, we assume that the system is configured such that the tags operate within the far-field range of the RFID reader's antenna, where $\| \mathbf{p}{_\text{R}}-\mathbf{p}{_\text{i}} \| \ge 2D^2/\lambda$ with $D$ being the maximum dimension of the RFID reader's antenna array~\cite{sherman1962properties}.

We utilize Alien AZ 9662 passive RFID tags, attaching two tags to the back of each hand. The setup employs an Impinj Speedway R420 RFID reader~\cite{impinj_r420} and two Vulcan RFID PAR90209H antenna with $9$ dBic antenna gain as well as elevation and azimuth beamwidth of $70^{\circ}$. The antennas are connected to a laptop equipped with ItemTest software, which enables communication between the reader and the RFID system.

The global coordinate system is aligned with the RFID reader’s, which employs a two-element uniform linear array with elements at $(-d/2,0)$ and $(d/2,0)$. The spacing between the elements is assumed to be $d = 0.8\lambda$, where $\lambda$ is the wavelength of the carrier's frequency. The RFID reader utilizes Smart Antenna Switching (SAS)~\cite{aiouaz2012rfid}. Initially, a signal is transmitted from the first antenna to detect the presence of tags within the antenna's field of view. If fewer than a predefined number of tags are detected\footnote{The reader software provides an option to set the upper limit on the number of tags that can be detected.}, the system switches to the second antenna. Otherwise, the tags found within the field of view of the first antenna are inventoried before switching to the second antenna. 




\subsection{Signal Model}
The passive RFID tag consists of an integrated circuit (IC) and an antenna, which is made up of two distinct segments forming a dipole. During reader-tag communication, the reader transmits a CW signal, and the tag's IC harvests power from the received signal. It then modulates the response by changing the antenna's load to backscatter or absorb the wave from the reader, resulting in the reflection coefficient, $\Gamma \in \{0,1\}$.\footnote{The reflection coefficient is calculated as : $\Gamma = \frac{Z_\text{Ant}-Z_\text{Load}}{Z_\text{Ant}+Z_\text{Load}}$, where $Z_\text{Ant}$ and $Z_\text{Load}$ denote the antenna impedance (typically around $50\Omega$) and the load impedance controlled by the tag's IC, respectively. When $Z_\text{Load}$ is matched with $Z_\text{Ant}$, i.e., $Z_\text{Load}= Z_\text{Ant}$, then  $\Gamma=0$, meaning that there is no reflection. If the $Z_\text{Load}=0$, i.e., a shortcut circuit, then $\Gamma=1$. In practice, due to imperfect load matching, the maximum value of the $\Gamma$ is close to $1$ and the minimum value is close to zero.}The RFID system is a SAS system and samples the backscattered signal from the tag(s) at each antenna element at different times, but periodically. 
The spatially sampled signals from each antenna are down-converted to a lower frequency using the oscillator and low-pass filter. These signals are subsequently fed into the analog-to-digital converter (ADC) to extract the in-phase and quadrature (IQ) components. The reader's system is based on a single transceiver card, meaning that a common oscillator is used for both transmission and reception. As a result, there will be no frequency offset in the baseband signal (IQ samples). 

 The AoA in the azimuth direction of the line-of-sight (LoS) path from the $i$-th tag relative to the center of the reader's antenna array is calculated as $\theta_i =   \text{atan2}\left( [\mathbf{q}_i]_2, [\mathbf{q}_i]_1 \right)- \pi/2$, where $\mathbf{q}_i = \frac{\mathbf{p}_i - \mathbf{p}_\text{R}}{\| \mathbf{p}_i - \mathbf{p}_\text{R} \|}$. 
 Similarly, the AoA of the $\ell$-th non-line-of-sight (NLoS) path associated with the $i$-th tag, which originates from the $\ell$-th scatterer located at $\mathbf{p}_{s_\ell}$, is computed as $\theta_i^\ell = \text{atan2}\left( [\mathbf{q}_i^\ell]_2, [\mathbf{q}_i^\ell]_1 \right)-\pi/2 $, where $\mathbf{q}_i^\ell = \frac{\mathbf{p}_{s_\ell} - \mathbf{p}_\text{R}}{\| \mathbf{p}_{s_\ell} - \mathbf{p}_\text{R} \|}$.
Therefore, under the above assumptions, the received signal at the $m$-th antenna of the reader, backscattered from the $i$-th tag, is denoted by $y_\text{m,i} (k) \in \mathbb{C}$ as:
\begin{align}\label{eq:ym}
    y_\text{m,i} (k) &= g_i \sqrt{P} \, a_\text{m}(\theta_\text{i}) \, s_\text{m}(k) 
    + \sum_{\ell = 1}^{L} g^{\ell}_i \sqrt{P} \, a_\text{m}(\theta^{\ell}_\text{i}) \, s_\text{m}(k) \nonumber \\ 
    &+ \nu_\text{m,i}(k), 
    \quad \forall m,i \in \{1, 2\}, \; \forall k \in \{1, 2, \dots, K\}
\end{align}
where $g_i$ and $g^{\ell}_i$ represent the channel gains of the round-trip path between the reader and the $i$-th tag for the LoS and the $\ell$-th NLoS paths, respectively. 
$s_\text{m}(k)= x(4k+ 2m+i-6)$ is the transmit signal over $m$-th antenna at time $k$, and $x(k)=e^{\jmath 2\pi f_\text{c} k T_\text{s} }$, $f_\text{c}$ is the the carrier frequency of the transmitted signal from the reader, $T_\text{s}$ is the sampling period of the ADC.  $\nu_\text{m,i}(k)$ indicates the effect of additive thermal noise at the reader,
 modeled as a zero-mean Gaussian distribution with variance $\sigma^2$. $P$ is the transmit power at the reader in the transmission mode. $L$ denotes the total number of NLoS paths. The $m$-th element of the steering vector is expressed as $a_\text{m}(\theta_\text{i})= \text{exp}(\jmath \frac{4\pi d}{\lambda } (m-1) \sin(\theta_\text{i}) ) \quad \forall m\in \{1,2\}$.\footnote{Unlike the standard steering vector mentioned in the literature, which is $a_\text{m}(\theta_\text{i})= \text{exp}(\jmath \frac{2\pi d}{\lambda } (m-1) \sin(\theta_\text{i}) ) \quad \forall m\in \{1,2\}$, the phase difference in the considered system is twice that of the one mentioned in the literature. This is due to the round-trip path between the reader and the tag.}
 We assume that the NLoS paths are weaker than the LoS path, which allows us to focus on the tag's AoA estimation in the subsequent step. Thus, we rewrite ~\eqref{eq:ym} as
\begin{align}\label{eq:ym_modified}
    y_\text{m,i} (k) &= g_i \sqrt{P} \, a_\text{m}(\theta_\text{i}) \, s_\text{m}(k) + \nu_\text{m,i}^{\prime}(k), \\ 
    &\quad \forall m,i \in \{1, 2\}, \; \forall k \in \{1, 2, \dots, K\}, \nonumber
\end{align}
where
$\nu_\text{m,i}^{\prime}(k) \triangleq \sum_{\ell = 1}^{L} g^{\ell}_i \sqrt{P} \, a_\text{m}(\theta^{\ell}_\text{i}) \, s_\text{m}(k) + \nu_\text{m,i}(k)$
represents the aggregated interference from NLoS paths and additive noise.

For the next step, we concatenate all the observations from each antenna element, as given in (\ref{eq:ym_modified}), over the observation time, i.e., $k\in \{1, \dots, K\}$. Thus, we have:
 
\begin{align}\label{eq:ym-vec}
    \mathbf{Y}_i = g_i \sqrt{P}  \boldsymbol{A}(\theta_\text{i}) \mathbf{S} + \boldsymbol{N}_i\quad \forall i \in \{1, 2\},  
\end{align}
where $\mathbf{Y}_i= \begin{bmatrix} \mathbf{y}_\text{1,i}^\top \\ \mathbf{y}_\text{2,i}^\top \end{bmatrix} \in \mathbb{C}^{2 \times K/4}$, $\boldsymbol{A}(\theta_\text{i})=\text{diag}(\boldsymbol{a}(\theta_\text{i}))  \in \mathbb{C}^{2\times2}$, $\boldsymbol{a}(\theta_\text{i})= [1\quad \text{exp}(\jmath \frac{4\pi d}{\lambda } (m-1) \sin(\theta_\text{i}) ) ]^\top $ $\mathbf{S} = \begin{bmatrix} \mathbf{s}_\text{1,i}^\top \\ \mathbf{s}_\text{2,i}^\top \end{bmatrix} \in \mathbb{C}^{2 \times K/4}$, where $\mathbf{s}_\text{1,i}$ and $\mathbf{s}_\text{2,i}$ are the transmit signals over the first and second antennas to the $i$-th tag, respectively. Also, $\mathbf{y}_\text{1,i} \in \mathbb{C}^{K/4}$ and $\mathbf{y}_\text{2,i} \in \mathbb{C}^{K/4}$ represent all the IQ samples from the first and second antenna elements from the $i$-th tag, respectively, over the observation time. Similarly, $\boldsymbol{N}_i = \begin{bmatrix} \boldsymbol{\nu^{\prime}}_\text{1,i}^\top \\ \boldsymbol{\nu^{\prime}}_\text{2,i}^\top \end{bmatrix} \in \mathbb{C}^{2 \times K/4}$, where $\boldsymbol{\nu^{\prime}}_\text{1,i} \in \mathbb{C}^{K/4}$ and $\boldsymbol{\nu^{\prime}}_\text{2,i} \in \mathbb{C}^{K/4}$ denote all the interference and noise samples from the first and second antenna elements during the reception from the $i$-th tag, respectively, over the observation time. 


\section{Problem formulation and Proposed Method}
In this section, we describe two main scenarios: first where one or two fixed tags are within the field of view of the readers, and another where the tag(s) are attached to the human body and move while performing gestures. In both scenarios, the goal is to estimate the AoA of the backscattered signal from the tags.

\subsection{Fixed Tag(s)}\label{Sec.Fixed Tag}
In this scenario, we first consider a single tag located at an unknown AoA with respect to the center of the reader's antenna. For multiple tags, the problem can be treated as independent single-tag AoA estimations since the reader's software provides separate IQ samples for each tag, as mentioned earlier. In some cases, a tag’s backscattered signal is detected by only one of the two antennas during part of the observation period. In such cases, the signal from the other antenna is discarded, because keeping it would make AoA estimation inaccurate.

\subsubsection{MUSIC Algorithm}\label{Sec. MUSIC}
For AoA estimation, we apply the MUltiple SIgnal Classification (MUSIC). After preprocessing the IQ samples and extracting the data corresponding to the target tag (in the case of multiple tags), we obtain the concatenated measurements as given in (\ref{eq:ym-vec}). Since the signal is wide-sense stationary (WSS) and satisfies the second-order ergodicity condition\footnote{
  $\mathbf{\hat{R}}_i \triangleq \mathbb{E}\{\mathbf{Y}_i \mathbf{Y}_i^\mathsf{H}\} \approx \frac{1}{K/4} \mathbf{Y}_i \mathbf{Y}_i^\mathsf{H}$
}, the covariance matrix $\mathbf{\hat{R}_i} \in \mathbb{C}^{2 \times 2}$ of $\mathbf{Y}_i$ is estimated as:

\begin{align}\label{eq:covar}
    \mathbf{\hat R}_i\triangleq \frac{1}{K/4} \ \mathbf{Y}_i \mathbf{Y}_i^{\mathsf{H}}\ \overset{(a)}{=} \alpha \boldsymbol{A}(\theta_\text{i}) \mathbf{R}^i_{\text{ss}} \boldsymbol{A}^{\mathsf{H}}(\theta_\text{i})+\mathbf{R}^i_{\text{n}} , 
\end{align}
where $\mathbf{R}^i_{\text{ss}} \triangleq \frac{1}{K/4} \mathbf{S}\mathbf{S}^{\mathsf{H}}$, 
$\alpha \triangleq g_i^2 P$, and $\mathbf{R}^i_{\text{n}}\triangleq \frac{1}{K/4} \boldsymbol{N}_i\boldsymbol{N}_i^{\mathsf{H}} $. (a) indicates that the noise and the signal are uncorrelated. Next, we perform eigen decomposition on $\mathbf{\hat R}_i$:
\begin{equation}\label{eq:decomposition}
    \mathbf{\hat R}_i\mathbf{u}_r={\lambda}_{r}\mathbf{u}_r, \quad \forall r\in\{n,s\},
\end{equation}
In which $\mathbf{u}_r$ is the eigenvector corresponding to the eigenvalue $\mathbf{\lambda}_r$, with $\mathbf{u}_{\text{n}} <\mathbf{u}_{\text{s}}$. Typically, eigenvectors associated with smaller eigenvalues ($ \mathbf{u}_{\text{n}}$) and larger eigenvalues ($ \mathbf{u}_{\text{s}}$) correspond to the noise subspace and signal subspace, respectively, which are orthogonal to each other. Considering these two orthogonal subspaces, we define the spectrum function, known as the MUSIC estimator, as follows~\cite[Eq. (6)]{schmidt1986multiple}:

\begin{equation}\label{eq:Music}
    P_{_\text{MUSIC}} (\theta_{\text{i}})=\frac{1}{\boldsymbol{a}^{\mathsf{H}}(\theta_{\text{i}}) \mathbf{u}_\text{n}  \mathbf{u}^{\mathsf{H}}_\text{n} \boldsymbol{a}(\theta_{\text{i}})}.
\end{equation}
Finally, we can find the peak of \( P_{\text{MUSIC}} \), which corresponds to the AoA of the \( i \)-th tag, by performing a line search over \( [\theta_{\text{min}}, \theta_{\text{max}}] \), where these values are defined based on our system's field of view. Indeed, our system's field of view is limited because the distance between the antennas is greater than \( \lambda/2 \), which leads to ambiguity in the AoA estimation. \footnote{Despite this limitation, the system setup is configured such that the AoA of the tag, resulting from gesture execution, falls within the unambiguous AoA range. }

\subsection{Moving tags attached to the Human Body}
In this scenario, we aim to track the AoA of the backscatter signals originating from moving tag(s) attached to the human body. To achieve this, similar to the previous section, we collect all IQ samples over the duration of the tags' movement. We then preprocess them, which includes extracting the IQ samples from each antenna for each tag, 
and discarding the samples measured at only one antenna for a period while the other antenna is missing, as discussed in the previous section.

Next, for each tag, we segment the pre-processed IQ samples into windows without overlap. We then estimate the AoA for each windowed IQ sample using the MUSIC algorithm, as explained in Sec.~\ref{Sec.Fixed Tag}. Hence, we have $\mathbf{z}= [\hat{\theta}_1,\hat{\theta}_2, \dots,\hat{\theta}_T]^\top$, where $T$ is the number of windows. 
However, some measurements are missing due to tag misdetection during gesture execution. Therefore, we apply a Kalman smoother to the AoAs estimated by MUSIC, leveraging the system dynamics. The smoother leverages future measurements to further refine past estimates.
We describe the Kalman smoother in the following sections.

Let $\boldsymbol{\theta}_t = [\theta^t_i, \omega^t_i]^\top$ represent the state of the dynamics of the changing AoA at $t$-th window (during the $i$-th tag movement), where $\omega^t_i$ is the rate of change of the AoA, assumed to be constant. Therefore, the state and measurement equations can be expressed as follows:

\begin{align}\label{eq:state}
    \boldsymbol{\theta}_t &= \mathbf{F} \boldsymbol{\theta}_{t-1} + \mathbf{w}_t, \\ \label{eq:measu}
    z_t &= \mathbf{H} \boldsymbol{\theta}_t + v_t, 
\end{align}
where $z_t\in \mathbb{R}$ denotes the measurement
at $t$-th window, $\mathbf{F} \triangleq \begin{bmatrix} \scriptstyle 1 & \scriptstyle \Delta t \\ \scriptstyle 0 & \scriptstyle 1 \end{bmatrix}$
 denotes the dynamic matrix, $\mathbf{H} \triangleq \begin{bmatrix} 1 & 0 \end{bmatrix}$ is the measurement matrix, and interval $\Delta t$ represents the time difference between two consecutive windows. $\mathbf{w}_t$ and $v_t$ are the process noise and measurement noise, respectively, which are assumed to follow a Gaussian distribution with zero mean and covariance $\mathbf{Q} = \begin{bmatrix} \scriptstyle \sigma^2_{\theta} & \scriptstyle 0 \\ \scriptstyle 0 & \scriptstyle \sigma^2_{\omega} \end{bmatrix}$
 and variance $\sigma^2_v$, and they are assumed to be independent. 
A Kalman smoother has the following steps:

\subsubsection{Prediction} In this step, the prior state estimate $\hat{\boldsymbol{\theta}}^{-}_t$ is obtained based on the considered dynamic system given in (\ref{eq:state}) as:
\begin{equation}\label{eq:predic}
    \hat{\boldsymbol{\theta}}^{-}_t= \mathbf{F}\hat{\boldsymbol{\theta}}_{t-1}.
\end{equation}
According to (\ref{eq:predic}), the covariance of the error of the predicted (prior) state estimate $\hat{\boldsymbol{\theta}}^{-}_t$, i.e., $\mathbf{P}^{-}_t$, can be calculated as:
\begin{equation}\label{eq:predic_cov_er}
    \mathbf{P}^{-}_t \triangleq \mathbb{E} \{e^{-}_t {e^{-}_t}^\top\} = \mathbf{F}\mathbf{P}_{t-1}\mathbf{F}^\top + \mathbf{Q},
\end{equation}
where $e^{-}_t \triangleq \boldsymbol{\theta}_t - \hat{\boldsymbol{\theta}}^{-}_t$, $\boldsymbol{\theta}_t$ is the true state at $t$-th window,  and $\mathbf{P}_{t-1}$ is the covariance of the error of the a posteriori state estimate $\hat{\boldsymbol{\theta}}_{t-1}$ at $(t-1)$-th window, which is explained in the next step. Note that we initialize the covariance of the predicted state error at the first window as $\mathbf{P}_0 = \mathbf{I}_2$.

\subsubsection{Measurements Update} In this step, we estimate the a posteriori $\hat{\boldsymbol{\theta}}_{t}$, based on the measurement available, $z_t$, provided by MUSIC in the current time window, and the prior state estimate, as follows:
\begin{equation}\label{eq:meas_update}
   \hat{\boldsymbol{\theta}}_t =  \hat{\boldsymbol{\theta}}^{-}_t + \mathbf{k}_t \left( z_t - \mathbf{H} \hat{\boldsymbol{\theta}}^{-}_t \right),
\end{equation}
where $\mathbf{k}_t \in \mathbb{R}^{2}$ is the Kalman gain in the $t$-th window, obtained by minimizing the mean squared a posteriori estimation error, i.e., the trace of the a posteriori error covariance matrix, $\mathbf{P}_t \triangleq \mathbb{E}\{ e_t e_t^\top \}$, with $e_t \triangleq \boldsymbol{\theta}_t - \hat{\boldsymbol{\theta}}_t$, as follows~\cite[Eq.~(15)]{sorenson1970least}:

\begin{equation}\label{eq:Kalman_gain}
\frac{d\,\mathrm{Tr}(\mathbf{P}_t)}{d \mathbf{k}_t} = 0 
\quad \Rightarrow \quad 
\mathbf{k}_t = \mathbf{P}^{-}_t \mathbf{H}^\top 
\bigl( \mathbf{H} \mathbf{P}^{-}_t \mathbf{H}^\top + \sigma_v^2 \bigr)^{-1},
\end{equation}

where $\mathbf{P}_t$ is updated as:
\begin{equation}\label{eq:Pt_update}
   \mathbf{P}_t = \left( \mathbf{I} - \mathbf{k}_t \mathbf{H} \right) \mathbf{P}^{-}_t.
\end{equation}
The above two steps (prediction and measurement update) are repeated sequentially until the estimate is updated based on the final AoA measurement corresponding to the last time window. Moreover, all prior and a posteriori state estimations, along with their covariance errors, are stored. This means that:
\begin{align} 
\label{eq:store_prio}
\hat{\boldsymbol{\Theta}}&= [\hat{\boldsymbol{\theta}}_1, \hat{\boldsymbol{\theta}}_2, \dots, \hat{\boldsymbol{\theta}}_T], \\ 
\label{eq:store_apost}
\hat{\boldsymbol{\Theta}}^{-}&= [\hat{\boldsymbol{\theta}}^{-}_1, \hat{\boldsymbol{\theta}}^{-}_2, \dots, \hat{\boldsymbol{\theta}}^{-}_T], \\ 
\label{eq:store_Cov_prio}
{\mathbf{P}}&= [{\mathbf{P}}_1, {\mathbf{P}}_2, \dots, {\mathbf{P}}_T], \\ 
\label{eq:store_Cov_apos}
{\mathbf{P}}^{-}&= [{\mathbf{P}}^{-}_1, {\mathbf{P}}^{-}_2, \dots, {\mathbf{P}}^{-}_T].
\end{align}
Note that, when no measurement is available due to tag misdetection, the \textit{Measurement Update} step will be skipped for that. 
\subsubsection{Smoother}
Next, we refine the above estimates by utilizing both future and past measurements through the Kalman filter's estimation. To achieve this, we perform the refinement in a backward direction, starting from the last window 
$T$. The backward recursion is given by ~\cite[Eq. (12.6)]{sarkka2023bayesian}:

\begin{align}\label{eq:smothgain}
    \mathbf{g}_t &= \mathbf{P}_t \mathbf{F}^\top [\mathbf{P}^{-}_{t+1}]^{-1}, \\ \label{eq:smoth state}
    \hat{\boldsymbol{\theta}}^s_t &= \hat{\boldsymbol{\theta}}_t +   \mathbf{g}_t \left ( \hat{\boldsymbol{\theta}}^s_{t+1} - \hat{\boldsymbol{\theta}}^{-}_{t+1} \right), \\ \label{eq:smoth cov}
    \mathbf{P}^s_t &= \mathbf{P}_t+ \mathbf{g}_t \left( 
 \mathbf{P}^s_{t+1}- \mathbf{P}^{-}_{t+1}\right)\mathbf{g}_t^\top , 
\end{align}
where $\mathbf{g}_t\in \mathbb{R}^{2}$ is the smoothing gain, $\hat{\boldsymbol{\theta}}^s_t$ is the smoothed state estimation, and $\mathbf{P}^s_t$ is the covariance error of the smoothed state estimate. We also initialize the smoothed state estimate and the covariance error of the smoothed state estimate as $\hat{\boldsymbol{\theta}}^s_T = \hat{\boldsymbol{\theta}}_T$ and $\mathbf{P}^s_T = \mathbf{P}_T$, respectively. 
Overall, the proposed method for AoA tracking can be summarized as described in \textit{Algorithm 1}.

\begin{algorithm}[!t]
\caption{AoA Tracking}
\begin{algorithmic}
\REQUIRE 
Received IQ samples, $\mathbf{Q}$, $\sigma^2_v$, $\mathbf{F}$, $\mathbf{P}_0$, $T$, $\theta_{\text{min}}$ and $\theta_{\text{max}}$
\STATE 1:Extract the number of tags, $N_t$, from the data.
\FOR{$i = 1$ to $N_t$}
\STATE \textbf{Pre-processing}
\STATE 2: Extract and store the IQ samples at each antenna from $i$-th tag. Discard the samples from one antenna when the other antenna’s samples are missing. 
\STATE \textbf{AoA Measurements}
\STATE 3: Segment the samples into $T$ non-overlapping windows.

\FOR{$t = 1$ to $T$}
    \STATE 4.1: Estimate the covariance matrix $\hat{\mathbf{R}}_i$ for the $t$-th window using (\ref{eq:covar}).
    \STATE 4.2: Calculate the noise subspace of $\hat{\mathbf{R}}_i$ using eigen decomposition given in \eqref{eq:decomposition}.
    \STATE 4.3: Estimate $\hat{\theta}_t$ by finding the peak of $P_{_\text{MUSIC}}$ given in ~\eqref{eq:Music} via a search over $\theta \in   [\theta_{\text{min}}, \theta_{\text{max}}]$.
\ENDFOR
\STATE 5: Store all the AoA measurements, i.e., $\mathbf{z}=[\hat{\theta}_1,\hat{\theta}_2, \dots,\hat{\theta}_T]^\top$.

\STATE \textbf{Kalman Filtering}
\FOR{$t = 1$ to $T$}
    \STATE 6.1 (Prediction): Estimate the prior state estimate $\hat{\boldsymbol{\theta}}^{-}_t$ and calculate ${\boldsymbol{P}}^{-}_t$ using ~\eqref{eq:predic} and ~\eqref{eq:predic_cov_er}, respectively.
    \STATE 6.2 (Measurements Update): \\ \quad Calculate the Kalman gain $\mathbf{k}_t$ using~\eqref{eq:Kalman_gain} \\ \quad Estimate a posteriori state $\hat{\boldsymbol{\theta}}_t$ using~\eqref {eq:meas_update} \\ \quad Calculate ${\boldsymbol{P}}_t$ using ~\eqref{eq:Pt_update}.
\ENDFOR
\STATE 7: Store all prior state and a posteriori state estimations, along with their covariance errors, i.e., $\hat{\boldsymbol{\Theta}}$, $\hat{\boldsymbol{\Theta}}^{-}$, ${\boldsymbol{P}}$, and ${\boldsymbol{P}}^-$ given in \eqref{eq:store_prio}--\eqref{eq:store_Cov_apos}. 

\STATE \textbf{Kalman Smoother}
\STATE 8: Set $\hat{\boldsymbol{\theta}}^s_T = \hat{\boldsymbol{\theta}}_T$ and $\mathbf{P}^s_T = \mathbf{P}_T$.
\FOR{$t = T$ to $1$}
    \STATE 9: Calculate the smoother gain $\mathbf{g}_t$ using ~\eqref{eq:smothgain}.
    \STATE 10: Estimate the smoothed state $\hat{\boldsymbol{\theta}}^s_t$ using ~\eqref{eq:smoth state}.
    \STATE 11: Calculate ${\boldsymbol{P}}^s_t$ using ~\eqref{eq:smoth cov}.
\ENDFOR 
\STATE 12: Store the AoA estimations for $i$-th tag, i.e., $\hat{\boldsymbol{\psi}}^s_i= \left[\hat{\boldsymbol{\theta}}^s_1[1], \hat{\boldsymbol{\theta}}^s_2[1], \dots, \hat{\boldsymbol{\theta}}^s_T[1]\right]^\top$ 
\ENDFOR 
\RETURN $\hat{\boldsymbol{\Theta}}^s= \left[\hat{\boldsymbol{\psi}}^s_1, \hat{\boldsymbol{\psi}}^s_2, \dots, \hat{\boldsymbol{\psi}}^s_{N_t} \right]$
\end{algorithmic}
\end{algorithm}

\section{Results and Discussion}
\begin{figure*}[!t]
\centering 
\begin{subfigure}[t]{0.32\textwidth}
    \centering
    \includegraphics[width=\linewidth]{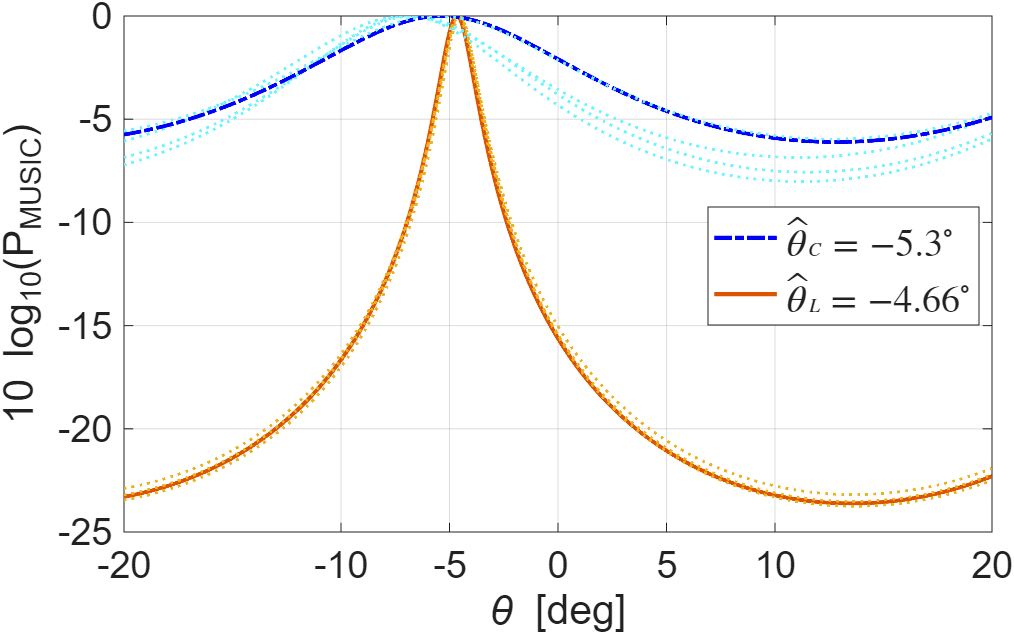}
    \caption{}
    \label{fig:Neg5}
\end{subfigure}
\hfill
\begin{subfigure}[t]{0.32\textwidth}
    \centering
    \includegraphics[width=\linewidth]{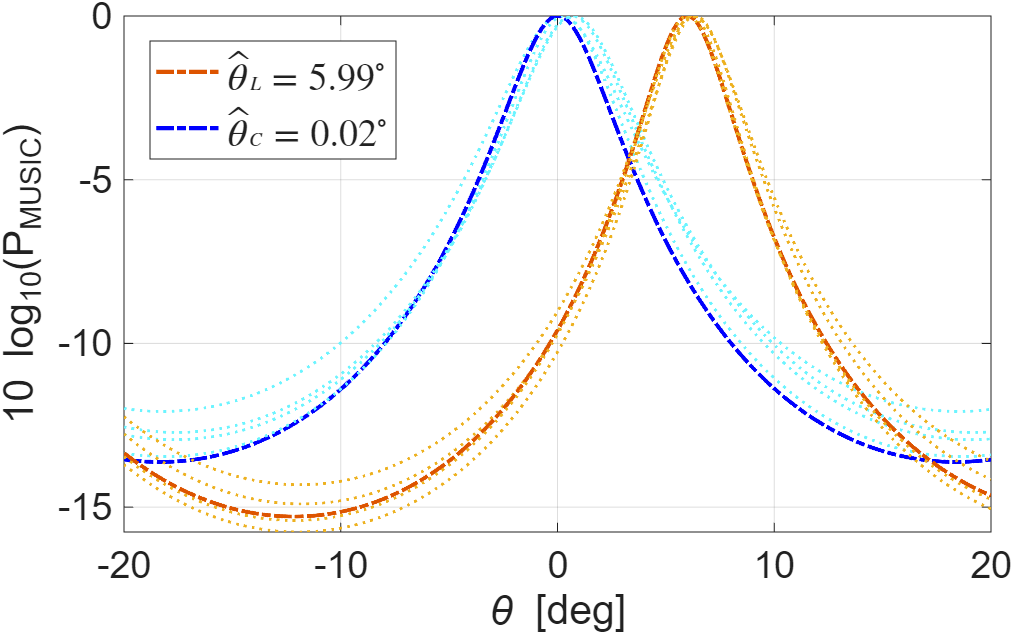}
    \caption{}
    \label{fig:Zero}
\end{subfigure}
\hfill
\begin{subfigure}[t]{0.32\textwidth}
    \centering
    \includegraphics[width=\linewidth]{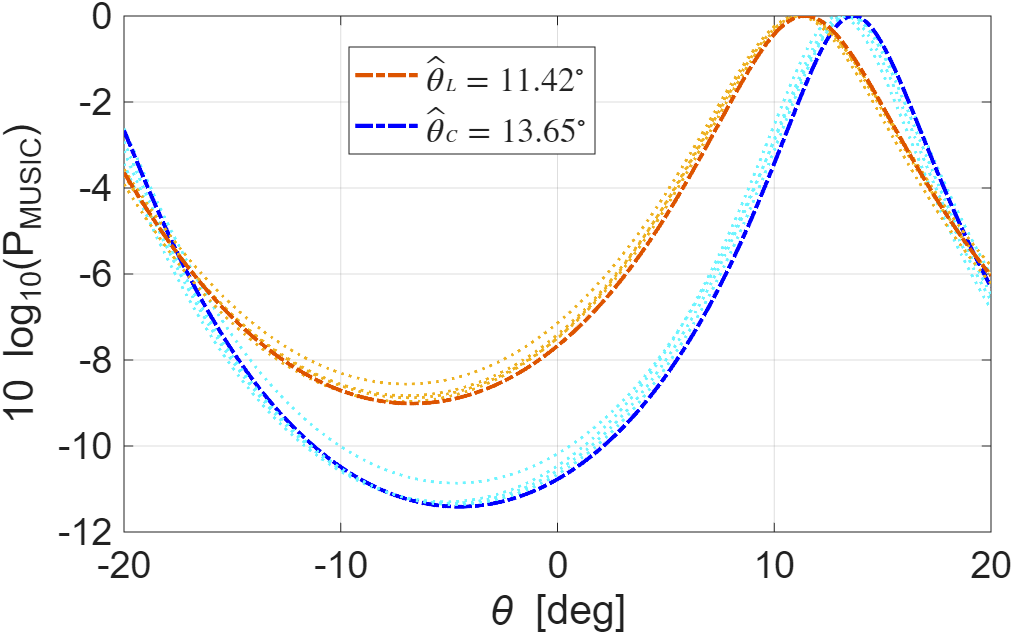}
    \caption{}
    \label{fig:Pos15}
\end{subfigure}
\caption{The evaluation of the MUSIC algorithm for AoA estimation is conducted for a fixed tag in the anechoic chamber ($\theta_{_\text{c}}$) and in the industrial laboratory ($\theta_\text{L}$) when the tag is located at: (a) $\theta = -5^\circ$, (b) $\theta = 0^\circ$, and (c) $\theta = 15^\circ$.
}
\label{fig:SingleTagAoA}
\end{figure*}
In this section, we evaluate the performance of the proposed algorithm. In our experiment, we set the transmit power at the reader to $P = 30~\text{dBm}$. The maximum sensitivity is set to $-84~\text{dBm}$. All these settings are configured via the RFID reader's software (see more details about the RFID reader's software in~\cite{impinj_connect}). The distance between the line where the tag(s) with an unknown AoA are placed and the line where the RFID antennas are located is $3~\text{m}$, fulfilling the far-field assumption. The reader transmits a signal with a carrier frequency of $f_c = 865.7~\text{MHz}$, corresponding to a wavelength of $\lambda = 34.65~\text{cm}$.

\subsection{AoA estimation for Fixed Tags}

In this section, we demonstrate that it is possible to estimate the AoA based on the data captured by an RFID reader. We consider two scenarios for our evaluation: one with a single tag and another with two tags. 

\subsubsection{One Tag}
In this case, we place the tag at different angles and compare the estimated and true values in two environments: the anechoic chamber and the industrial laboratory, as shown in Fig.~\ref{fig:SingleTagAoA}. This figure presents five experiments for each environment, highlighting the best result. As observed, the estimation error in the industrial laboratory is larger compared to that in the anechoic chamber due to the multipath effect. For example, as shown in Fig.~\ref{fig:Pos15}, when the tag is located at $\theta = 15^\circ$, the AoA estimation error is approximately $1.35^\circ$ for the anechoic chamber and $3.58^\circ$ for the industrial laboratory. For $\theta = 0^\circ$, as shown in Fig.~\ref{fig:Zero}, the AoA estimation error in the industrial laboratory is significantly higher, approximately $6^\circ$. The main reason is the presence of a strong non-line-of-sight path around $0^\circ$. Note that some errors arise from the setup process, including aligning the RFID reader's antenna and positioning the tags precisely. However, the results demonstrate that the AoA estimator performs reasonably well for AoA tracking in gesture recognition, which will be discussed later.
\subsubsection{Two Tags}
\begin{figure}
\centering
\includegraphics[width=0.7\linewidth]{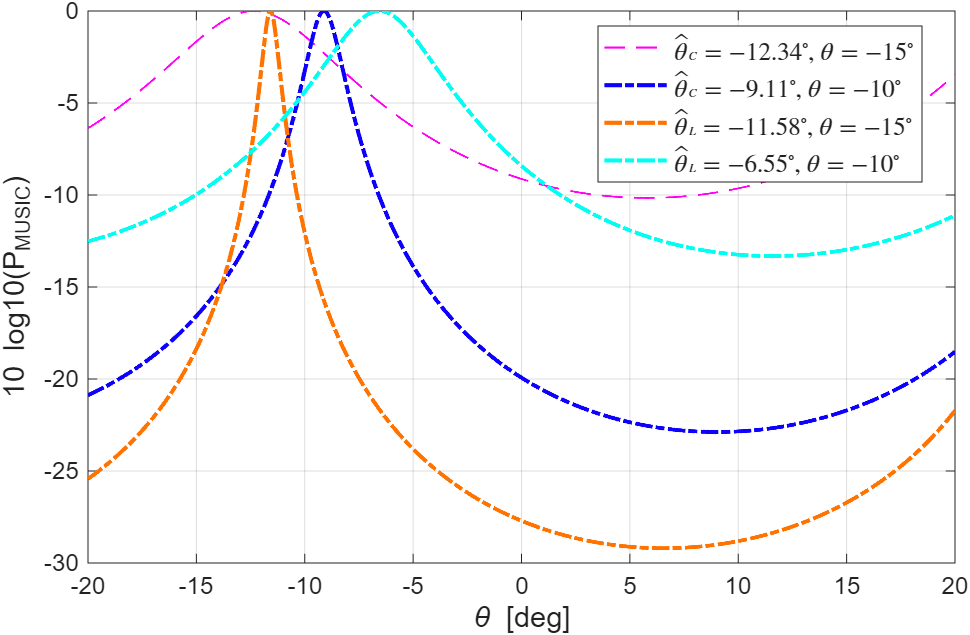}
\caption{The evaluation of AoA estimation with two tags, located at $-15^\circ$ and $-10^\circ$, in the anechoic chamber and the industrial laboratory.}
\label{fig:twofixedTags}
\end{figure}

Fig.~\ref{fig:twofixedTags} demonstrates the AoA estimation where there are two tags at different angles backscattering signals. As observed, the AoA estimation error in the industrial laboratory is still acceptable compared to that in the anechoic chamber (for example, at angle $-10^\circ$, the error is $3.45^\circ$ in the industrial laboratory with the presence of multipath, compared to the error of $0.9^\circ$ in the anechoic chamber). This validates the AoA measurements for tracking in the case of multiple tags.

\subsection{AoA tracking as a complementary feature for gesture differentiation}
\begin{figure*}
\begin{subfigure}{0.125\textwidth}
    \includegraphics[width=\linewidth]{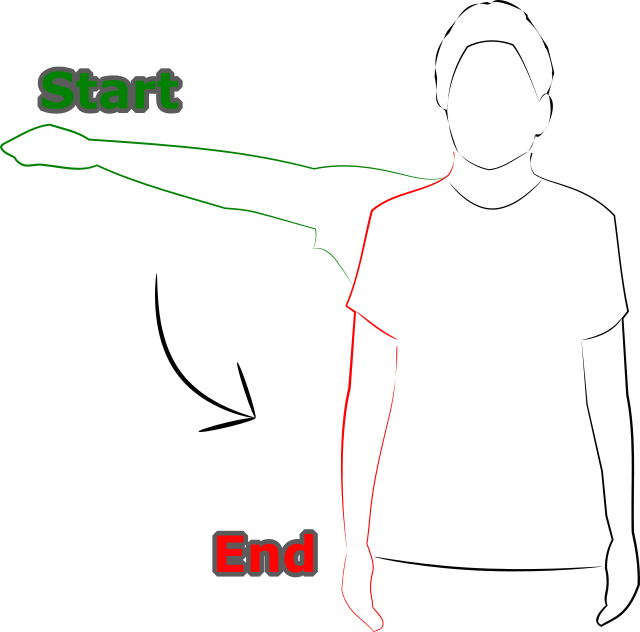}
    \caption{}
    \label{g}
\end{subfigure}
\begin{subfigure}{0.12\textwidth}
    \includegraphics[width=\linewidth]{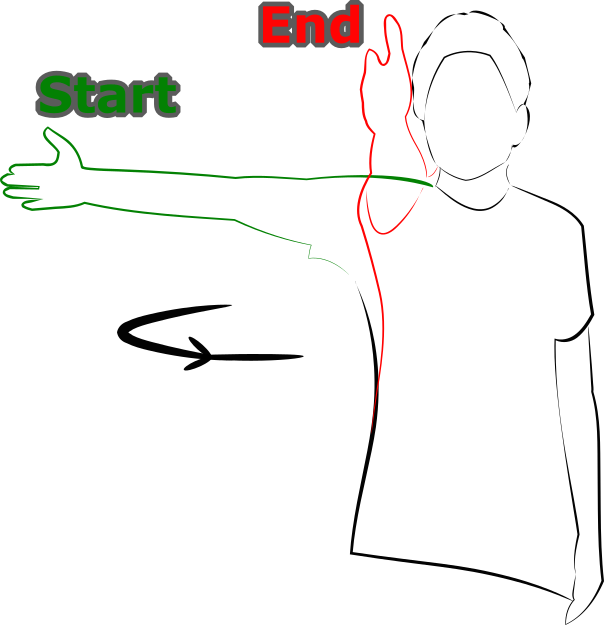}
    \caption{}
    \label{g}
\end{subfigure}
\begin{subfigure}{0.125\textwidth}
    \includegraphics[width=\linewidth]{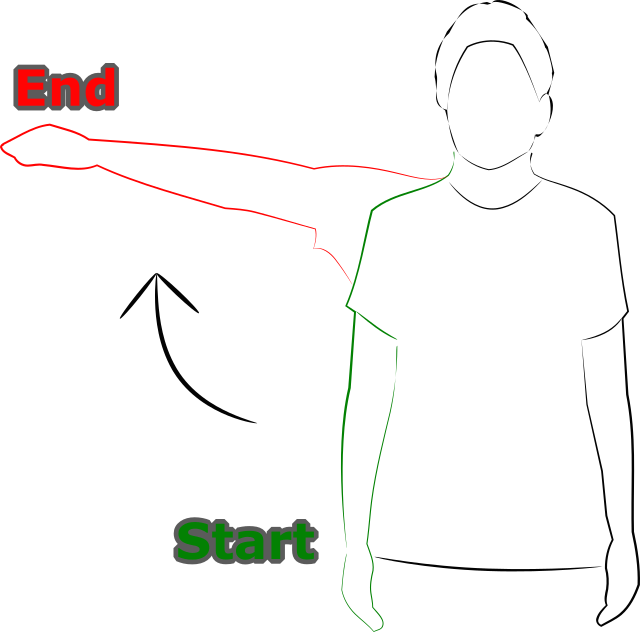}
    \caption{}
    \label{g}
\end{subfigure}
\begin{subfigure}{0.09\textwidth}
    \includegraphics[width=\linewidth]{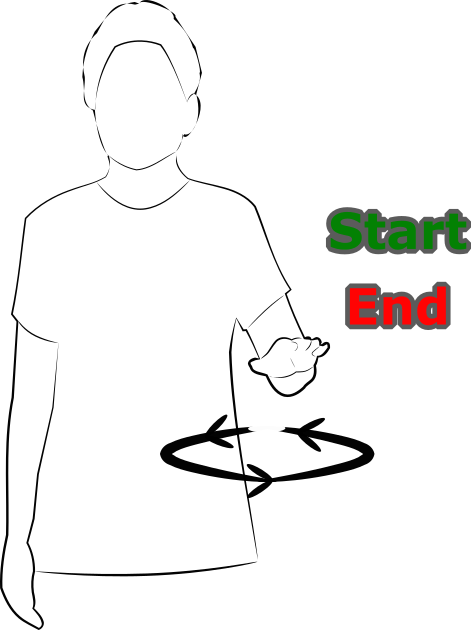}
    \caption{}
    \label{v}
\end{subfigure}
\begin{subfigure}{0.095\textwidth}
    \includegraphics[width=\linewidth]{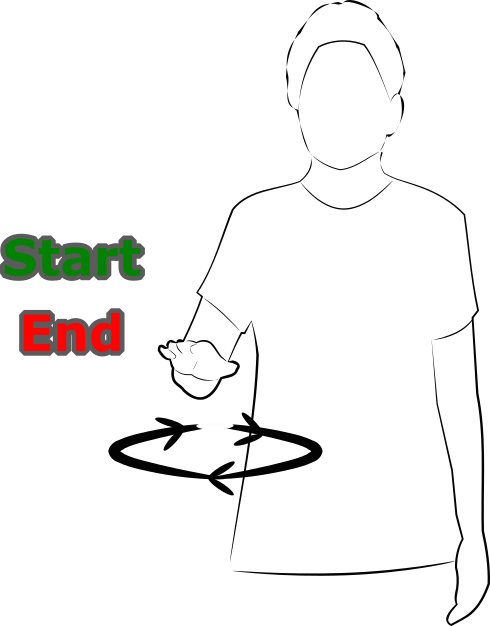}
    \caption{}
    \label{v}
\end{subfigure}
\begin{subfigure}{0.085\textwidth}
    \includegraphics[width=\linewidth]{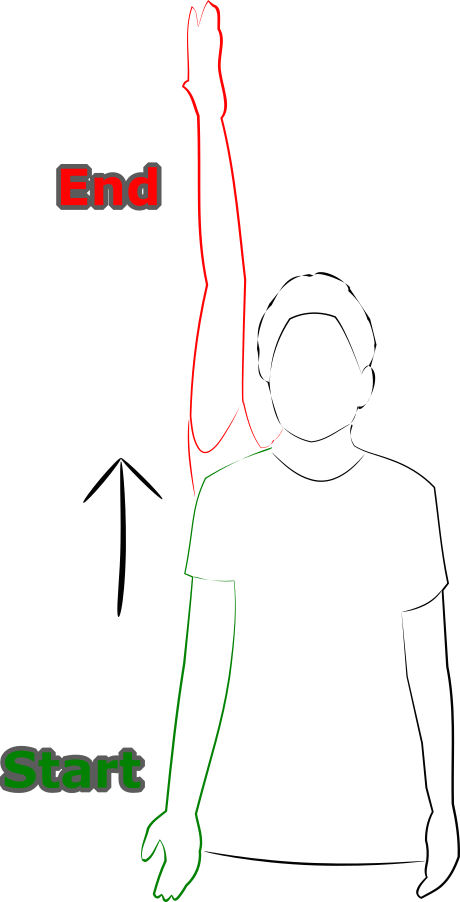}
    \caption{}
    \label{v}
\end{subfigure}
\begin{subfigure}{0.085\textwidth}
    \includegraphics[width=\linewidth]{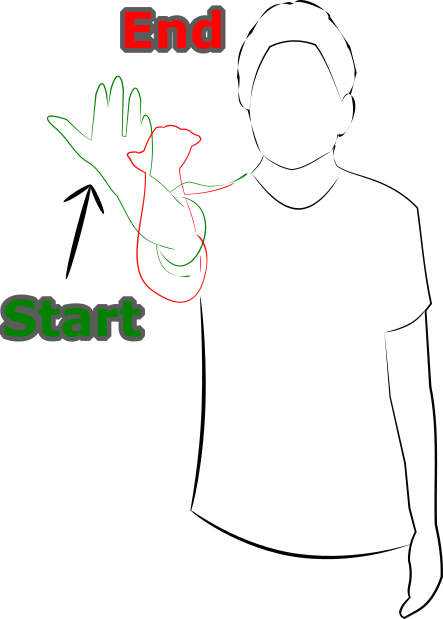}
    \caption{}
    \label{v}
\end{subfigure}
\begin{subfigure}{0.095\textwidth}
    \includegraphics[width=\linewidth]{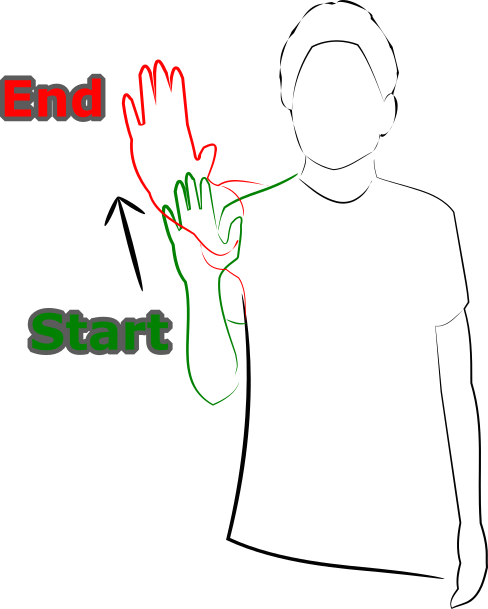}
    \caption{}
    \label{v}
\end{subfigure}
\begin{subfigure}{0.145\textwidth}
    \includegraphics[width=\linewidth]{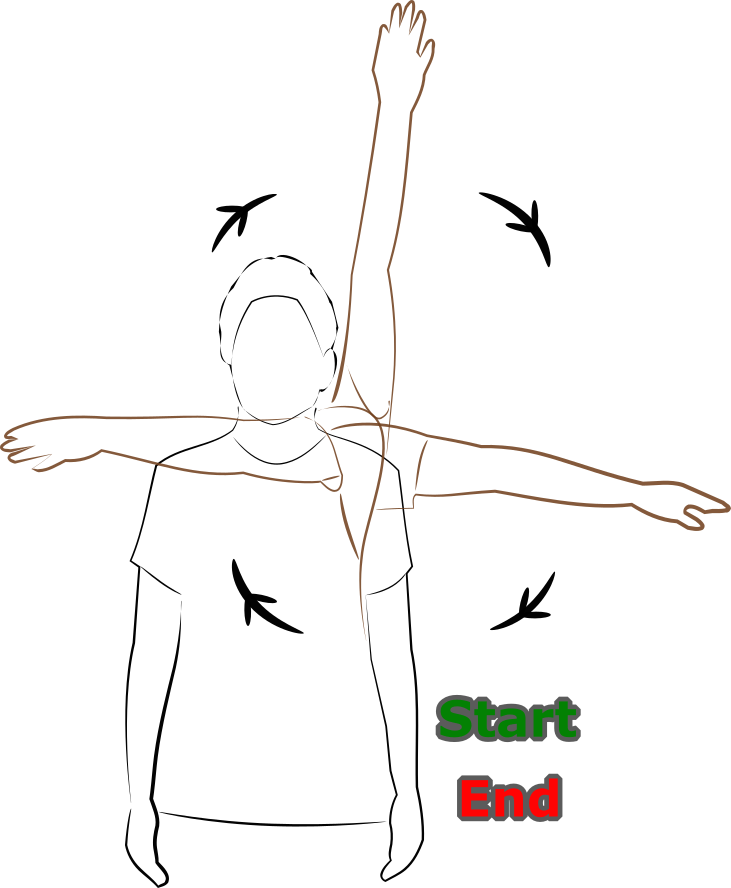}
    \caption{}
    \label{v}
\end{subfigure}
\begin{subfigure}{0.145\textwidth}
    \includegraphics[width=\linewidth]{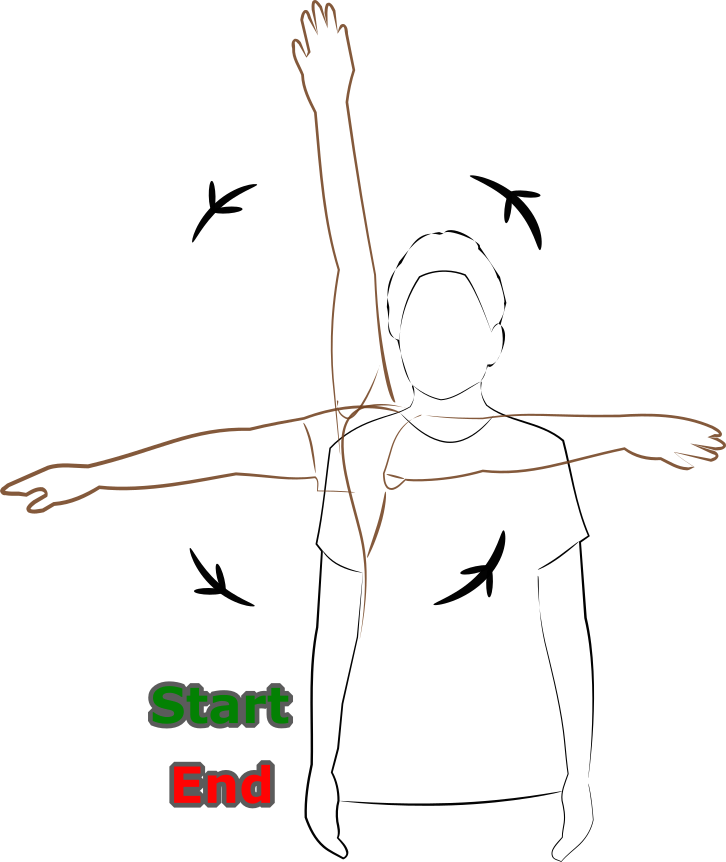}
    \caption{}
    \label{v}
\end{subfigure}
\begin{subfigure}{0.133\textwidth}
    \includegraphics[width=\linewidth]{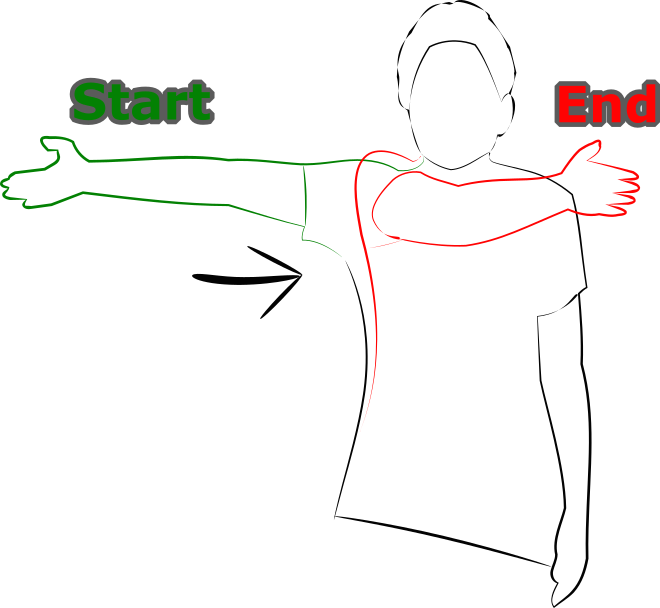}
    \caption{}
    \label{v}
\end{subfigure}
\begin{subfigure}{0.138\textwidth}
    \includegraphics[width=\linewidth]{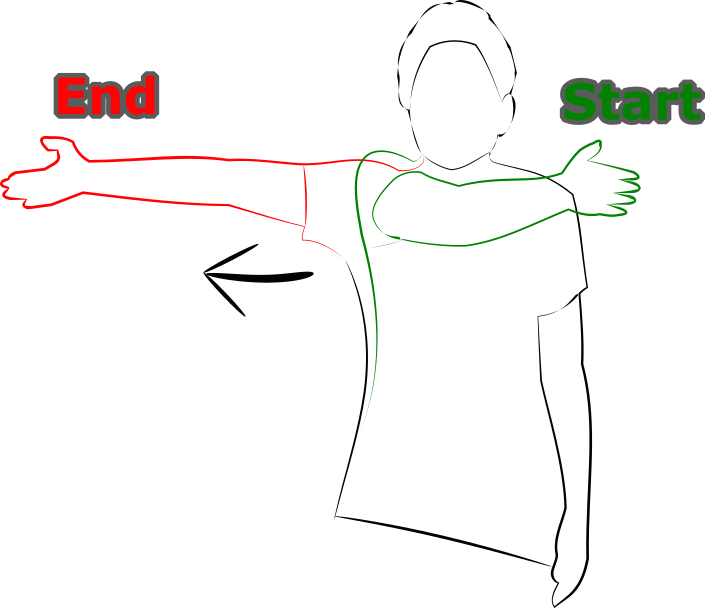}
    \caption{}
    \label{v}
\end{subfigure}
\begin{subfigure}{0.183\textwidth}
    \includegraphics[width=\linewidth]{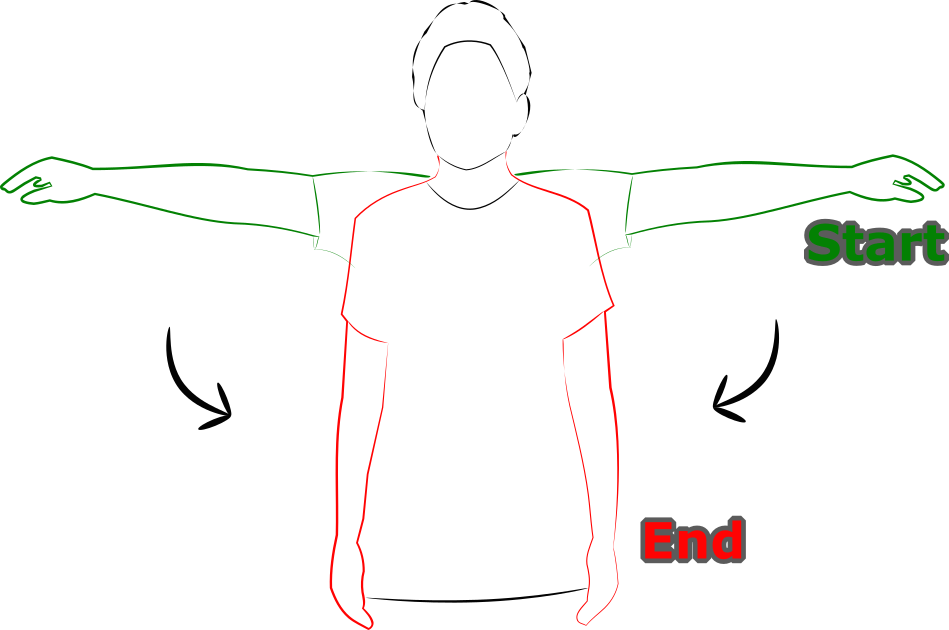}
    \caption{}
    \label{v}
\end{subfigure}
\begin{subfigure}{0.188\textwidth}
    \includegraphics[width=\linewidth]{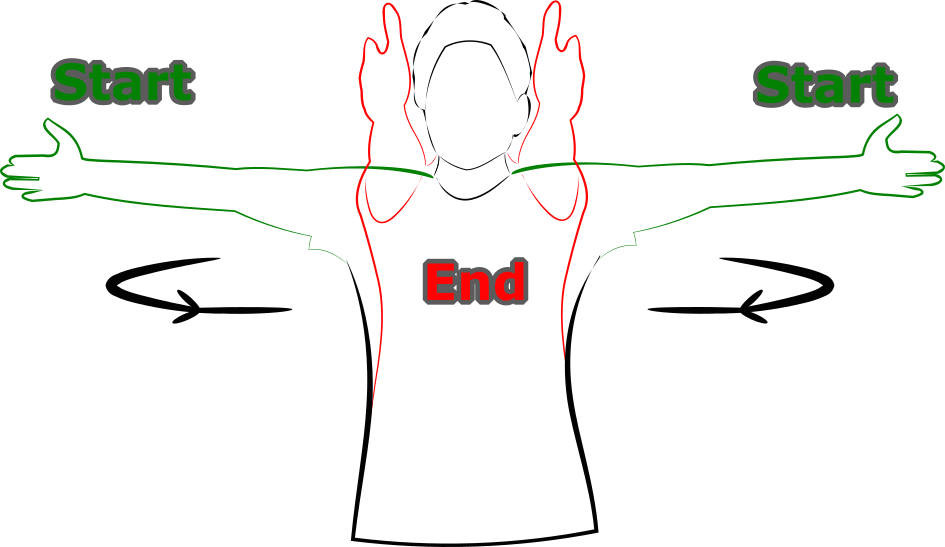}
    \caption{}
    \label{v}
\end{subfigure}
\begin{subfigure}{0.19\textwidth}
    \includegraphics[width=\linewidth]{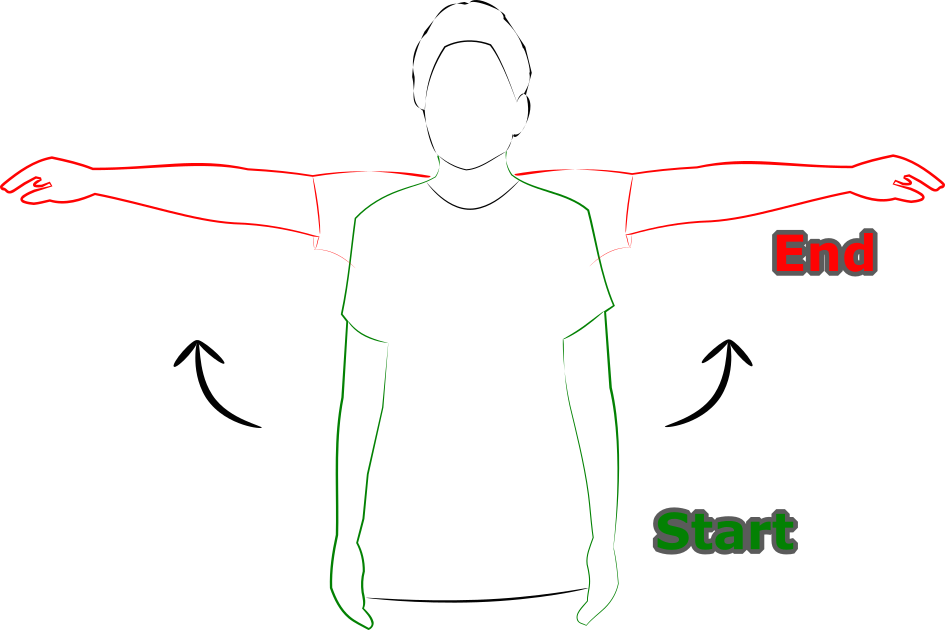}
    \caption{}
    \label{v}
\end{subfigure}
\begin{subfigure}{0.14\textwidth}
    \includegraphics[width=\linewidth]{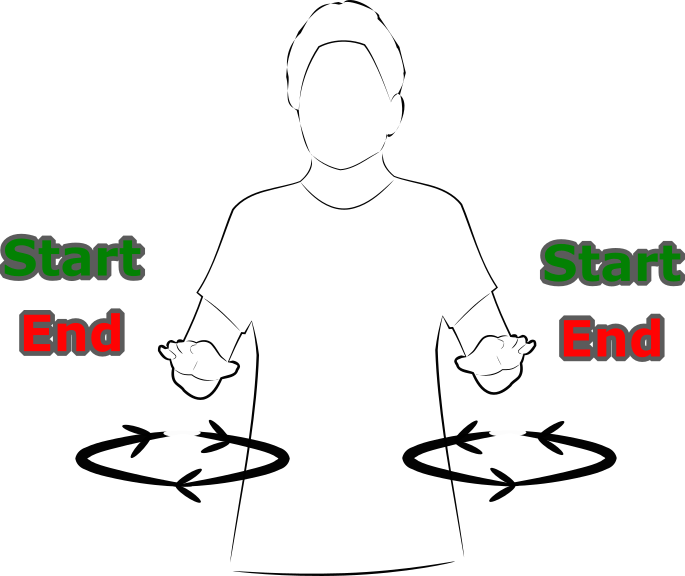}
    \caption{}
    \label{v}
\end{subfigure}
\begin{subfigure}{0.14\textwidth}
    \includegraphics[width=\linewidth]{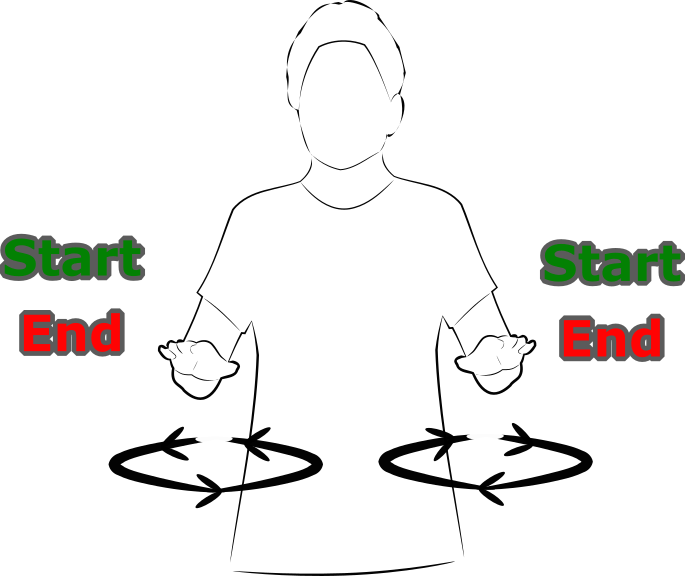}
    \caption{}
    \label{v}
\end{subfigure}
\begin{subfigure}{0.09\textwidth}
    \includegraphics[width=\linewidth]{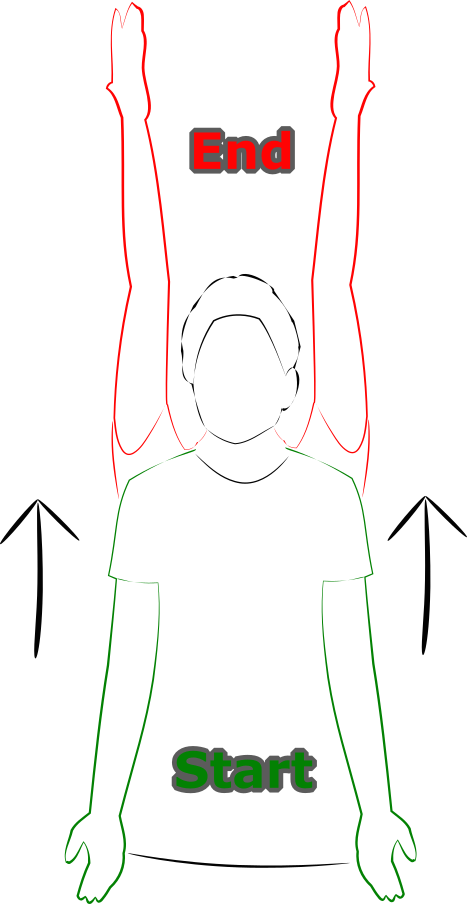}
    \caption{}
    \label{v}
\end{subfigure}
\begin{subfigure}{0.1\textwidth}
    \includegraphics[width=\linewidth]{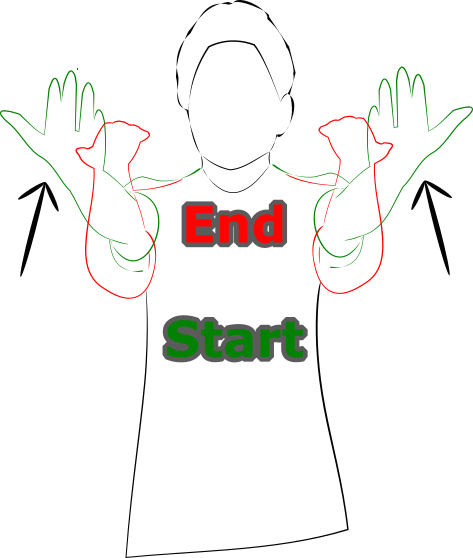}
    \caption{}
    \label{v}
\end{subfigure}
\begin{subfigure}{0.1\textwidth}
    \includegraphics[width=\linewidth]{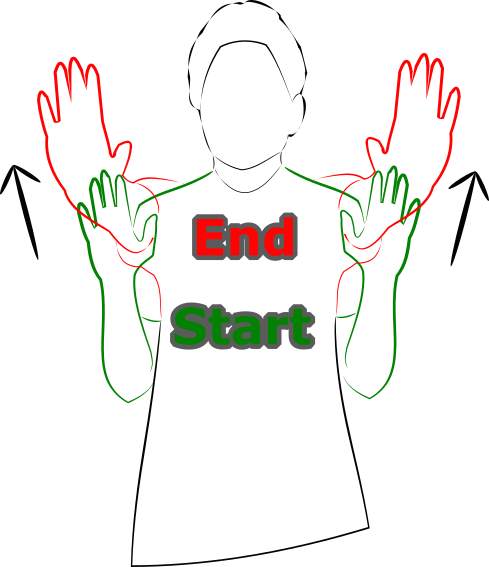}
    \caption{}
    \label{v}
\end{subfigure}
\begin{subfigure}{0.12\textwidth}
    \includegraphics[width=\linewidth]{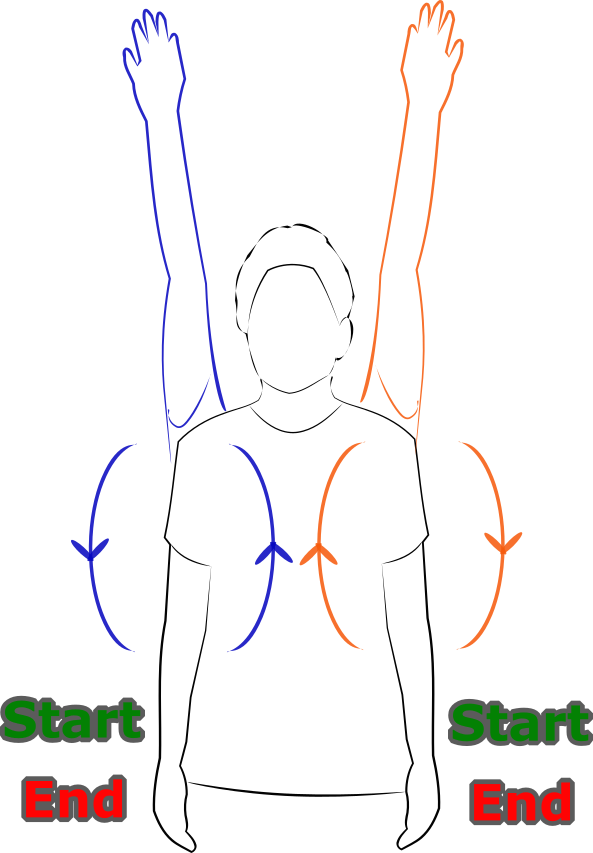}
    \caption{}
    \label{v}
\end{subfigure}
\caption{The gestures performed by participants are (a) 
LD: Lateral Down (1), (b) LF: Lateral to Front (2), (c) LR: Lateral Raise (3), (d) LAC: Left Arm Circle (4), (e) RAC: Right Arm Circle (5), (f) L: Lift (6), (g) Pl: Pull (7), (h) Ps: Push (8), (i) LRo: Left Round (9), (j) RR: Right Round (10), (k) SL: Swipe Left (11), (l) SR: Swipe Right (12), (m) 2HLD: Two Hands Lateral Down (13), (n) 2HLF: Two Hands Lateral to Front (14), (o) 2HLR: Two Hands Lateral Raise (15), (p) 2HIC: Two Hands Inward Circle (16), (q) 2HOC: Two Hands Outward Circle (17), (r) 2HL: Two Hands Lift (18), (s) 2HPl: Two Hands Pull (19), (t) 2HPs: Two Hands Push (20), (u) 2HR: Two Hands Round (21).}
\label{Gestures}
\end{figure*}

For data collection, participants were positioned 3 meters away from the antennas.
The experiment included 5 participants (2 males, 3 females) with diverse body types and heights ranging from 155 cm to 185 cm. Each participant repeated every gesture 20 times. With a total of 21 unique gestures, the resulting dataset consisted of 2100 samples. As illustrated in Fig.~\ref{Gestures}, the gestures included simple one-handed and complex two-handed movements.

\begin{figure*}[!t]
\centering 
\begin{subfigure}[t]{0.24\textwidth}
    \centering
    \includegraphics[width=\linewidth]{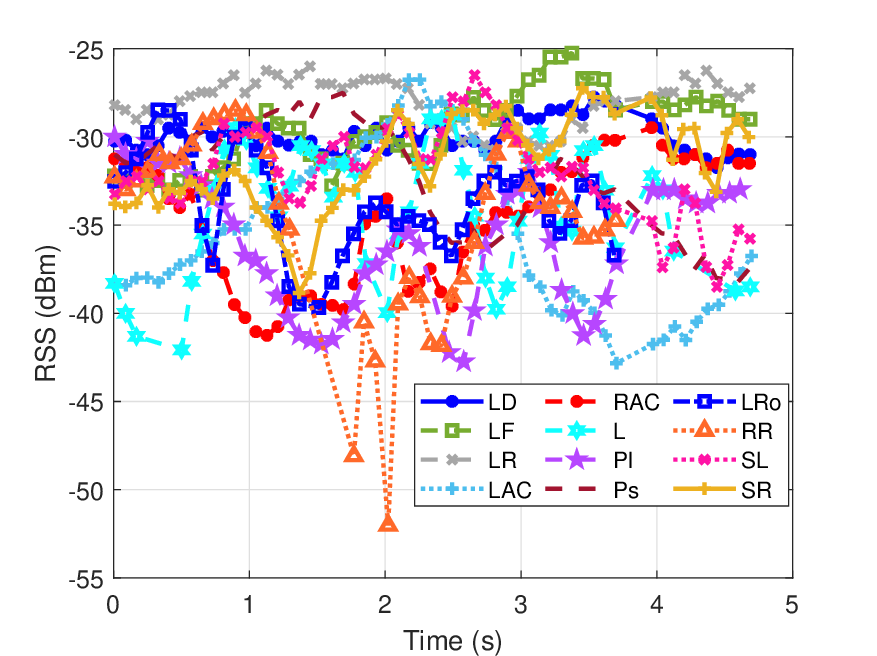}
    \caption{}
    \label{RssGestures}
\end{subfigure}
\hfill
\begin{subfigure}[t]{0.24\textwidth}
    \centering
    \includegraphics[width=\linewidth]{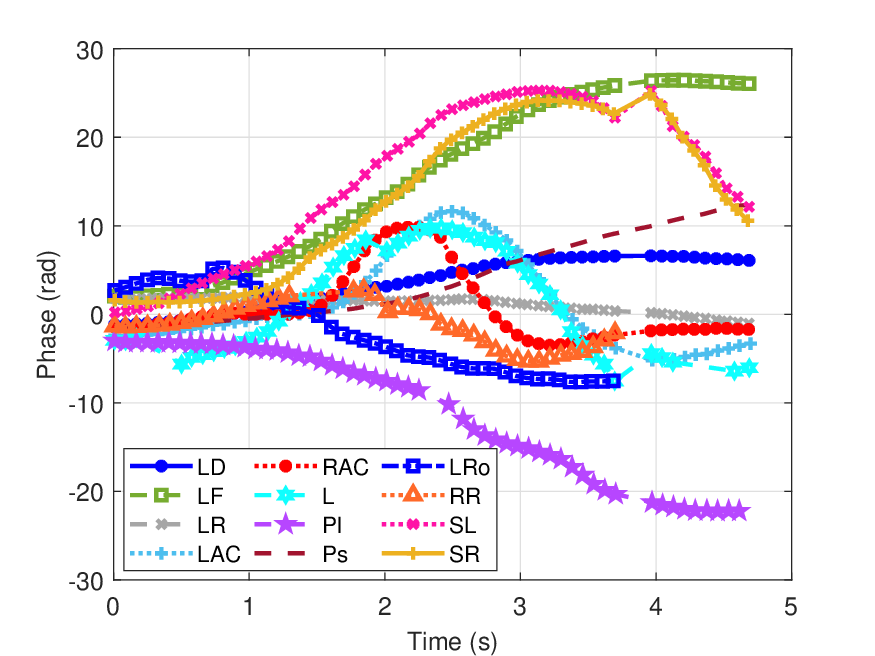}
    \caption{}
    \label{PhaseGestures}
\end{subfigure}
\hfill
\begin{subfigure}[t]{0.24\textwidth}
    \centering
    \includegraphics[width=\linewidth]{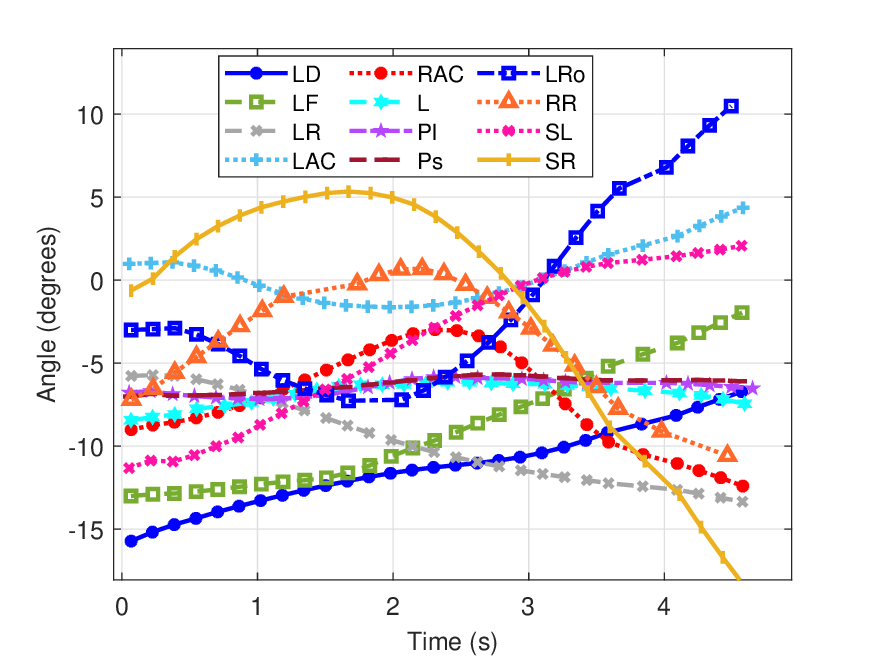}
    \caption{}
    \label{AoAGestures}
\end{subfigure}
\hfill
\begin{subfigure}[t]{0.24\textwidth}
    \centering
    \includegraphics[width=\linewidth]{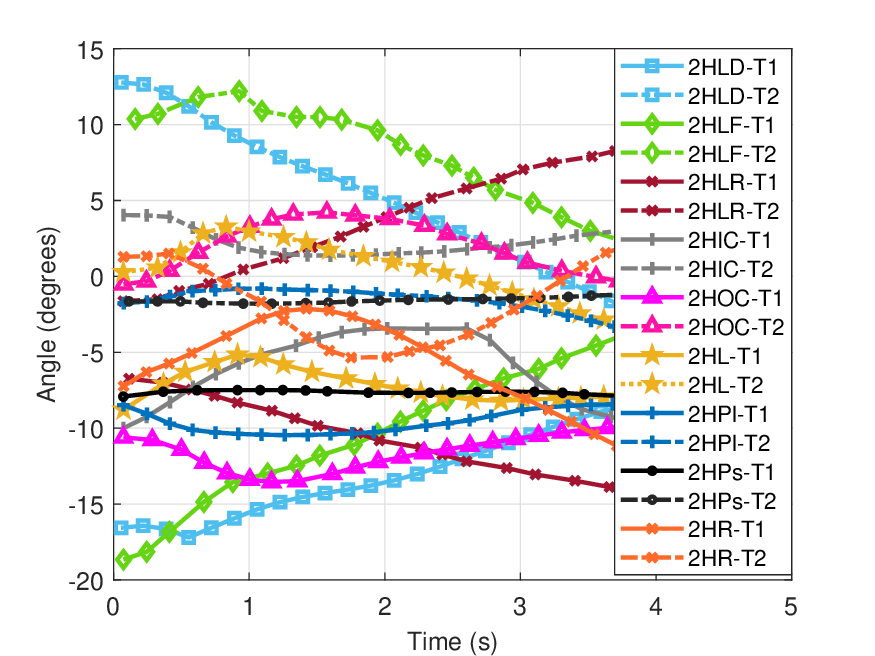}
    \caption{}
    \label{AoAComplexGestures}
\end{subfigure}
\caption{
Comparison of RSS, phase, and AoA features for gesture differentiation. 
(a) Processed RSS signals from the hand-mounted tag for a simple set of gestures. 
(b) Processed phase signals from the hand-mounted tag for a simple set of gestures. (c) Estimated AoA signals from the hand-mounted tag for a simple set of gestures. (d) Estimated AoA signals from the hand-mounted tags for a complex set of gestures.
T1 and T2 denote Tag~1 and Tag~2, respectively.
}
\label{RSSPhaseAoAComparison}
\end{figure*}

Fig.~\ref{RSSPhaseAoAComparison} illustrates the advantages of AoA tracking in a gesture recognition system. Fig.~\ref{RssGestures} presents the RSS values extracted for the simple one-handed gesture set. RSS is known to be highly sensitive to environmental factors such as multipath fading, shadowing, and interference. As shown, the RSS signals for different gestures are largely indistinguishable, indicating that RSS alone is not a reliable metric for robust gesture recognition.
In contrast, phase information is generally considered more reliable in RF sensing applications, as it captures fine-grained characteristics of wave propagation, including timing and path differences. However, our experimental results indicate that the phase of backscattered signals from the tags often exhibits similarities across different gestures, as demonstrated in Fig.~\ref{PhaseGestures}. For instance, while the phase signals for the SL and SR gestures appear quite similar, their AoA estimations are markedly different, as shown in Fig.~\ref{AoAGestures}. Specifically, the estimated AoA for the SL begins in the negative range and moves toward the positive, whereas for SR, it starts near zero, shifts briefly toward the positive, and then trends negative—consistent with the expected motion of these two gestures.
Another example can be seen with the L and RAC gestures, which are not sufficiently distinguishable based on phase alone. However, when AoA is taken into account, these gestures become clearly separable.
Fig.~\ref{AoAComplexGestures} displays the estimated AoA from both tags for the complex two-handed gesture set. As an illustration, the 2HLR and 2HLD gestures exhibit AoA tracks in opposite directions, despite the gestures appearing similar in execution. Likewise, the 2HIC and 2HOC gestures produce distinct AoA signal trends, even though they are performed similarly.
This analysis confirms that, while RSS and phase may lack the discriminatory power required for accurate gesture recognition, AoA tracking provides spatial information that enhances recognition performance. Accordingly, the following section presents a comprehensive evaluation demonstrating the effectiveness of AoA features, both individually and when integrated with RSS and phase data, across various recognition models.

\subsection{Gesture recognition using tracked AoAs}
In this section, we evaluate AoA tracking in gesture recognition and demonstrate how this metric enhances the performance of various models compared to using only RSS and phase signal. 

We applied benchmark approaches.
We implemented Reactor~\cite{zhang2022real}, which extracts features from the signal and trains a Random Forest Classifier (RFC). We also employed a Support Vector Machine (SVM)~\cite{cortes1995support} as an alternative classifier. For each feature configuration, RFC and SVM models were trained independently. For brevity, we denote them using RFC/SVM, referring to two separate models per configuration. Based on different feature extraction strategies from the available metrics, RSS, phase, and AoA, we constructed models as detailed below:
\begin{itemize}
\item\textbf{RFC/SVM with SP} utilizes only statistical features extracted from the phase.
\item\textbf{RFC/SVM with SWP} combines statistical features and wavelet coefficients derived from the phase.
\item\textbf{RFC/SVM with SPR} includes statistical features extracted from both the phase and RSS.
\item\textbf{RFC/SVM with SA} uses only statistical features extracted from AoA.
\item\textbf{RFC/SVM with SWA} combines statistical features and wavelet coefficients derived from AoA.
\item\textbf{RFC/SVM with SPRA} incorporates statistical features extracted from the phase, RSS, and AoA.

\end{itemize}

We extract statistical features from signals, i.e., RSS, phase, and AoA, including the mode, the median, the first quartile, the third quartile, the mean, the max, the min, the range, the variance, the standard deviation, the third-order central moment, the kurtosis, the skewness, and the entropy. On top of that, Pearson correlations are extracted between every pair of RSS, phase, and AoA data corresponding to the two tags.
Moreover, we use discrete wavelet transform with Daubechies wavelet and extract low-frequency coefficients and concatenate them to statistical features.

In addition, AoA tracking was evaluated using other benchmark models, including the Early Fusion approach presented in~\cite{calatrava2023light}, in which each tag’s RSS and phase from both antennas, along with the AoAs corresponding to the two tags, are fed in parallel to its network. The Late Fusion method described in~\cite{golipoor2024rfid} and the EUIGR method~\cite{yu2019rfid} are also applied, in which features from the signals are first extracted through their respective networks and then merged. Additionally, GRfid~\cite{zou2016grfid}, which is a dynamic time warping-based algorithm, is utilized for either phase or AoA separately.

The performance metrics, including accuracy, precision, recall, and F1-score, are presented in TABLE.~\ref{tab1} for models that use phase or  phase combined with RSS as input features, and in TABLE.~\ref{tab2} for models that use AoA or its combination with phase and RSS as input features. In general, incorporating AoA leads to a notable improvement in all metrics across all methods. For example, the accuracy of the RFC model using statistical features from phase and RSS is 85.49\% in TABLE.~\ref{tab1}. However, when statistical features from AoA are added, i.e., RFC with SPRA, the accuracy increases substantially to 97.2\%. Similarly, the SVM-based model achieves 78.11\% without AoA, while concatenating AoA features increases the model’s accuracy to 93.12\%. 

In contrast with Early Fusion in TABLE.~\ref{tab1}, which achieves an accuracy of 83.46\%, the same approach in TABLE.~\ref{tab2} (with AoA) shows a considerable improvement, reaching 96.94\%. A similar trend is observed with Late Fusion, where the accuracy increases from 83.96\% to 95.92\%. While the EUIGR method shows an accuracy of 80.66\%, it achieves 95.41\% when AoA is incorporated. GRfid also benefits from AoA, with its performance increasing by almost 12\%.

\begin{figure*}[htbp]
\centering
\begin{subfigure}{0.19\textwidth}
    \includegraphics[width=\linewidth]{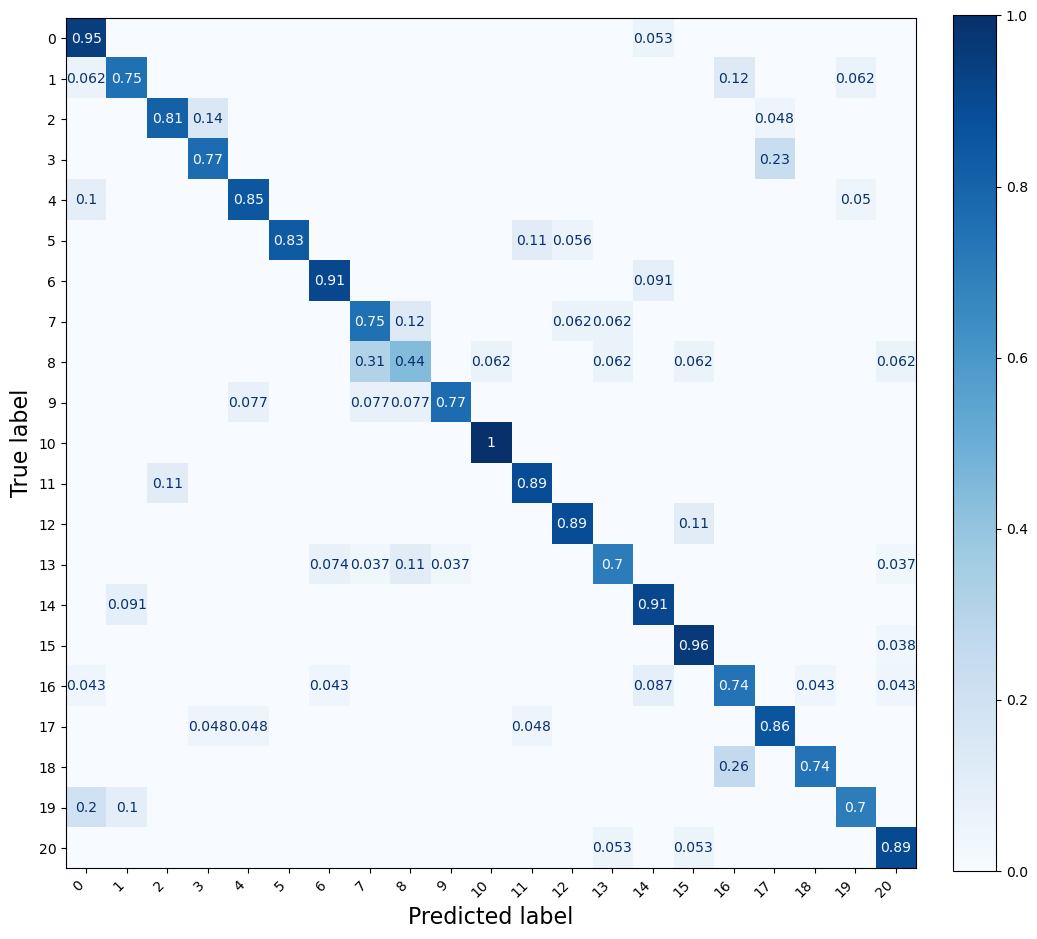}
    \caption{Accuracy=82.18\%}
    \label{Conf:sub1}
\end{subfigure}
\hfill
\begin{subfigure}{0.19\textwidth}
    \includegraphics[width=\linewidth]{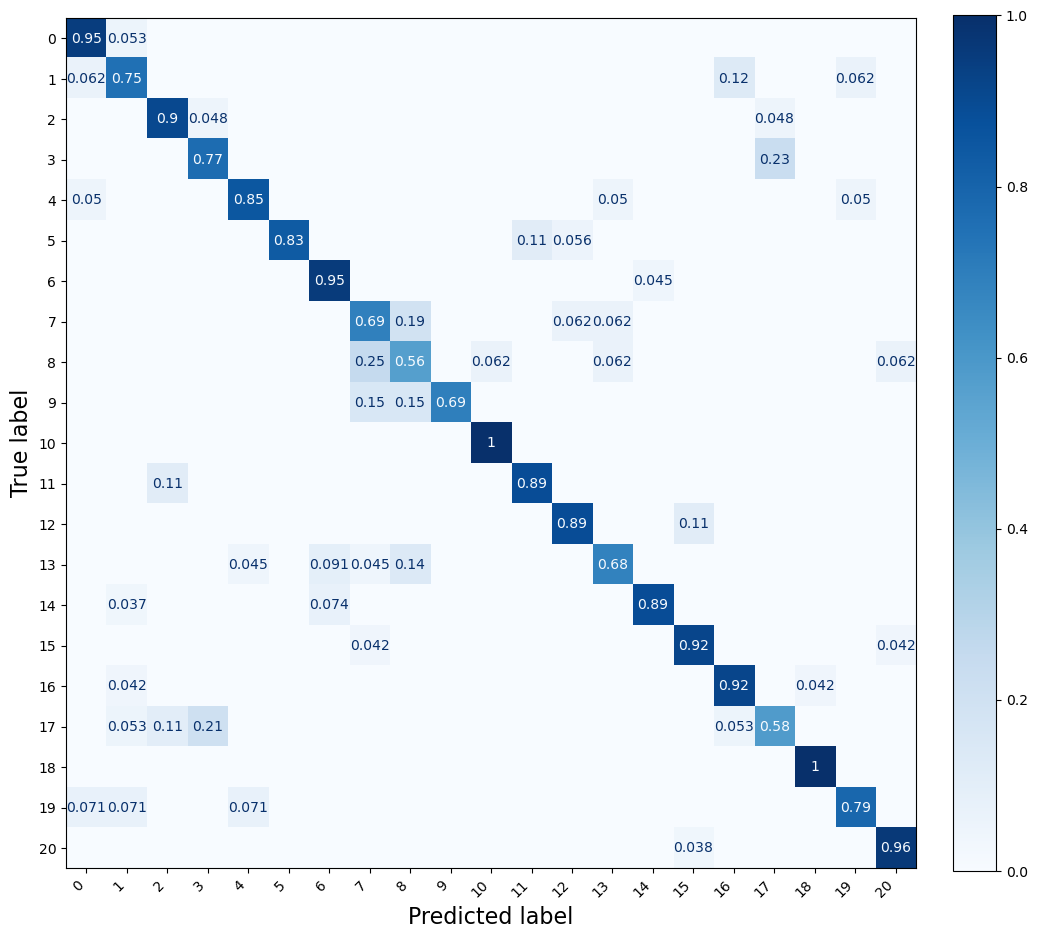}
    \caption{Accuracy=84.22\%}
    \label{Conf:sub2}
\end{subfigure}
\hfill
\begin{subfigure}{0.19\textwidth}
    \includegraphics[width=\linewidth]{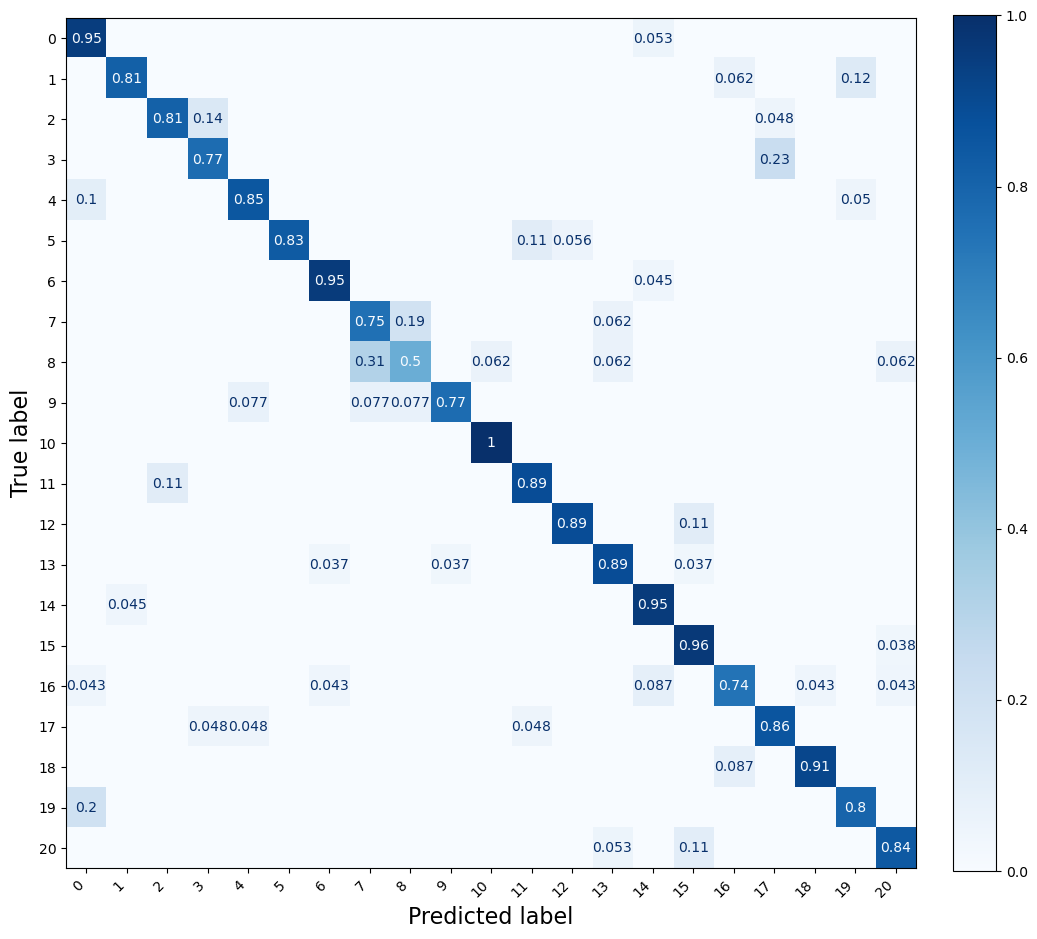}
    \caption{Accuracy=85.49\%}
    \label{Conf:sub3}
\end{subfigure}
\hfill
\begin{subfigure}{0.19\textwidth}
    \includegraphics[width=\linewidth]{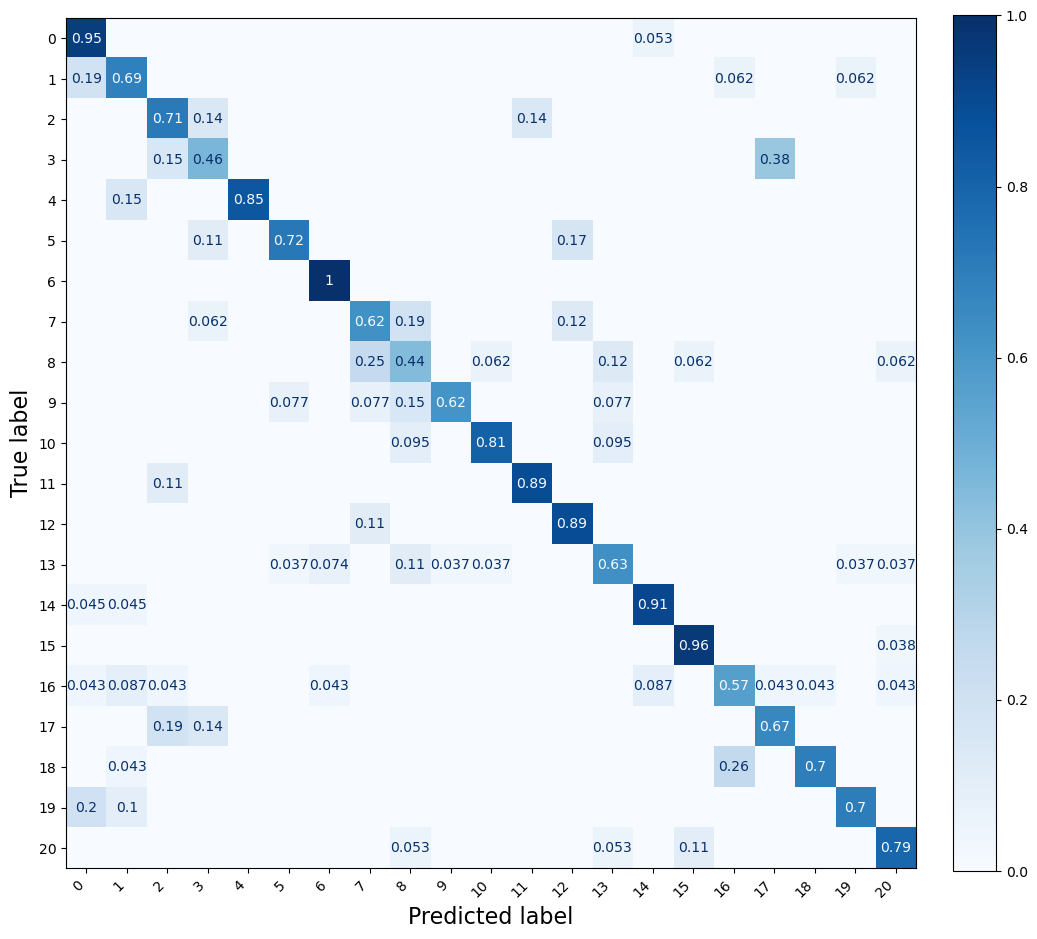}
    \caption{Accuracy=75.06\%}
    \label{Conf:sub4}
\end{subfigure}
\hfill
\begin{subfigure}{0.19\textwidth}
    \includegraphics[width=\linewidth]{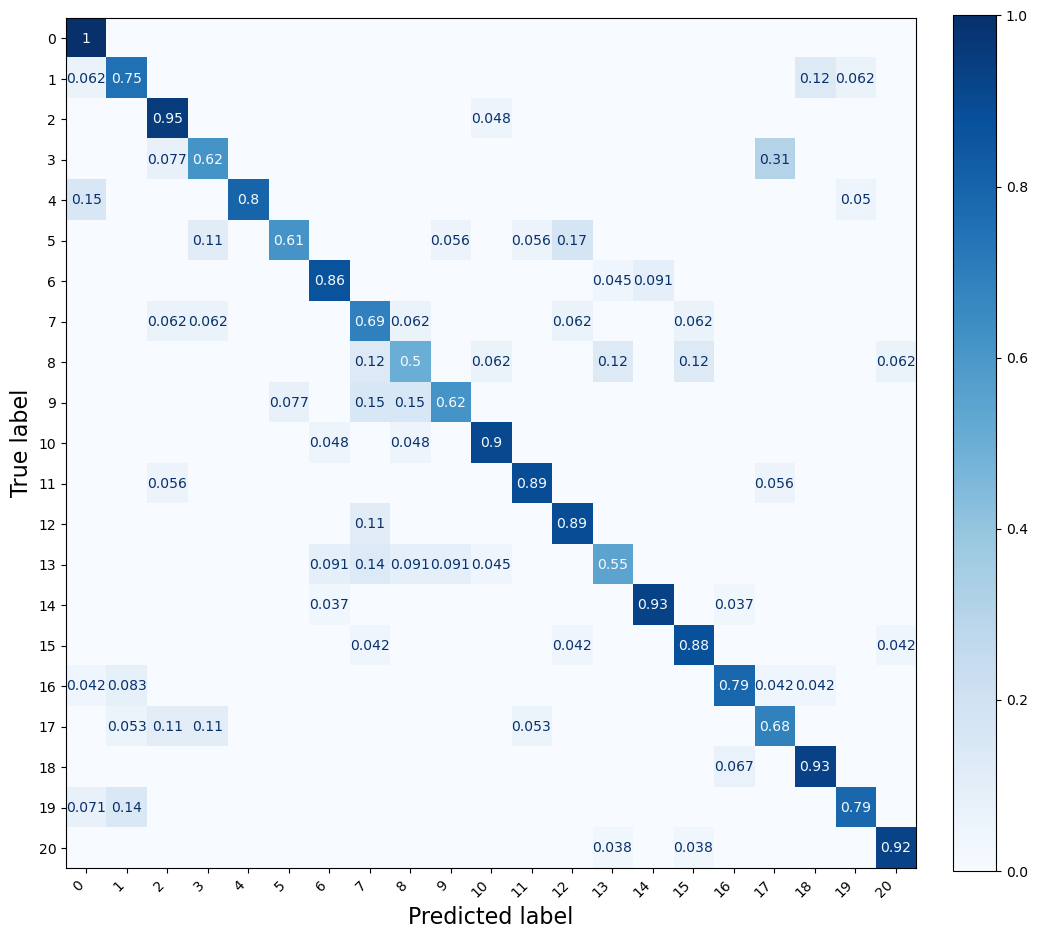}
    \caption{Accuracy=79.89\%}
    \label{Conf:sub5}
\end{subfigure}

\begin{subfigure}{0.19\textwidth}
    \includegraphics[width=\linewidth]{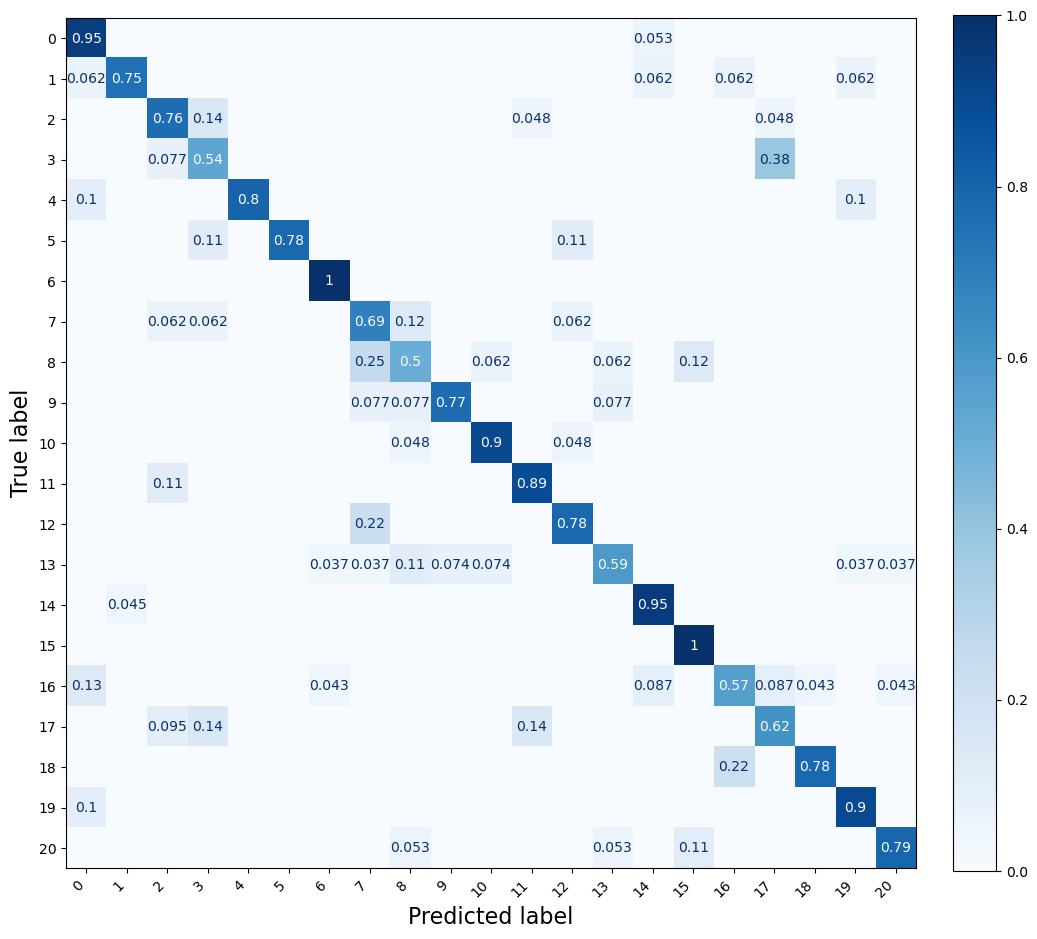}
    \caption{Accuracy=78.11\%}
    \label{Conf:sub6}
\end{subfigure}
\hfill
\begin{subfigure}{0.19\textwidth}
    \includegraphics[width=\linewidth]{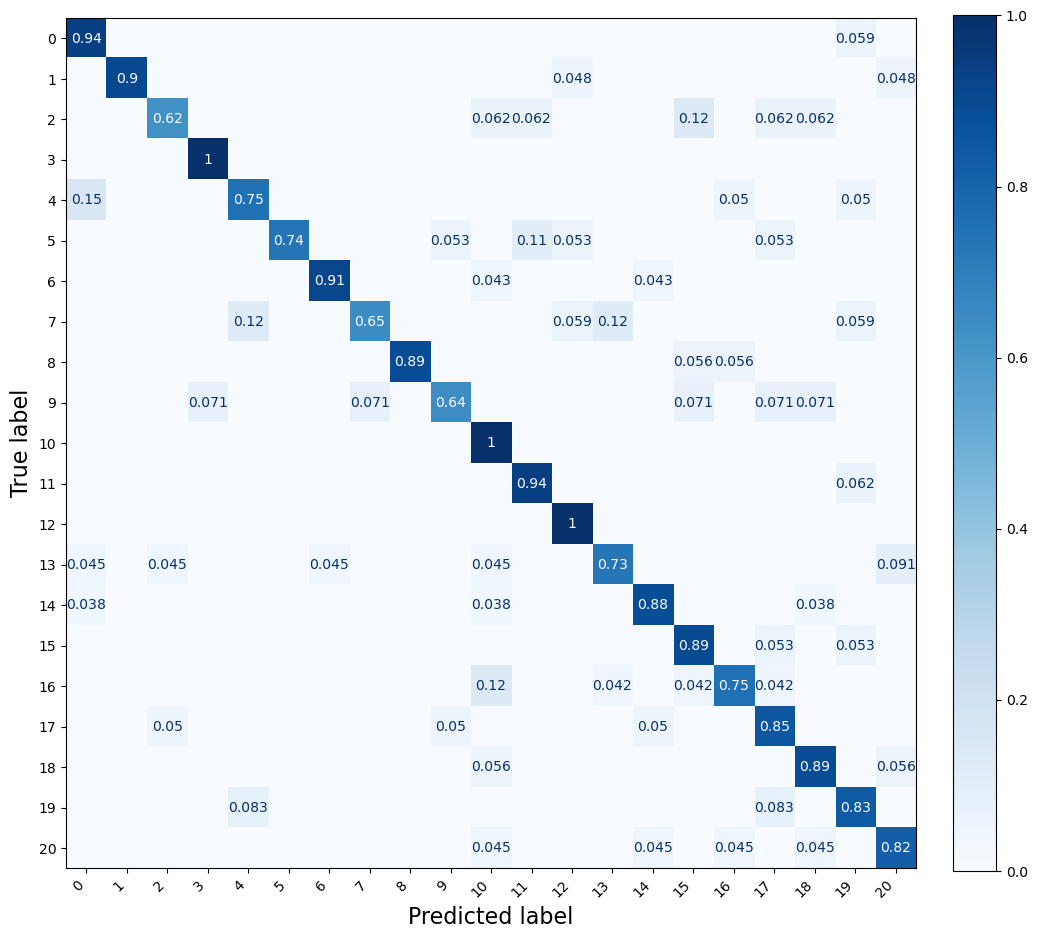}
    \caption{Accuracy=83.96\%}
    \label{Conf:sub7}
\end{subfigure}
\hfill
\begin{subfigure}{0.19\textwidth}
    \includegraphics[width=\linewidth]{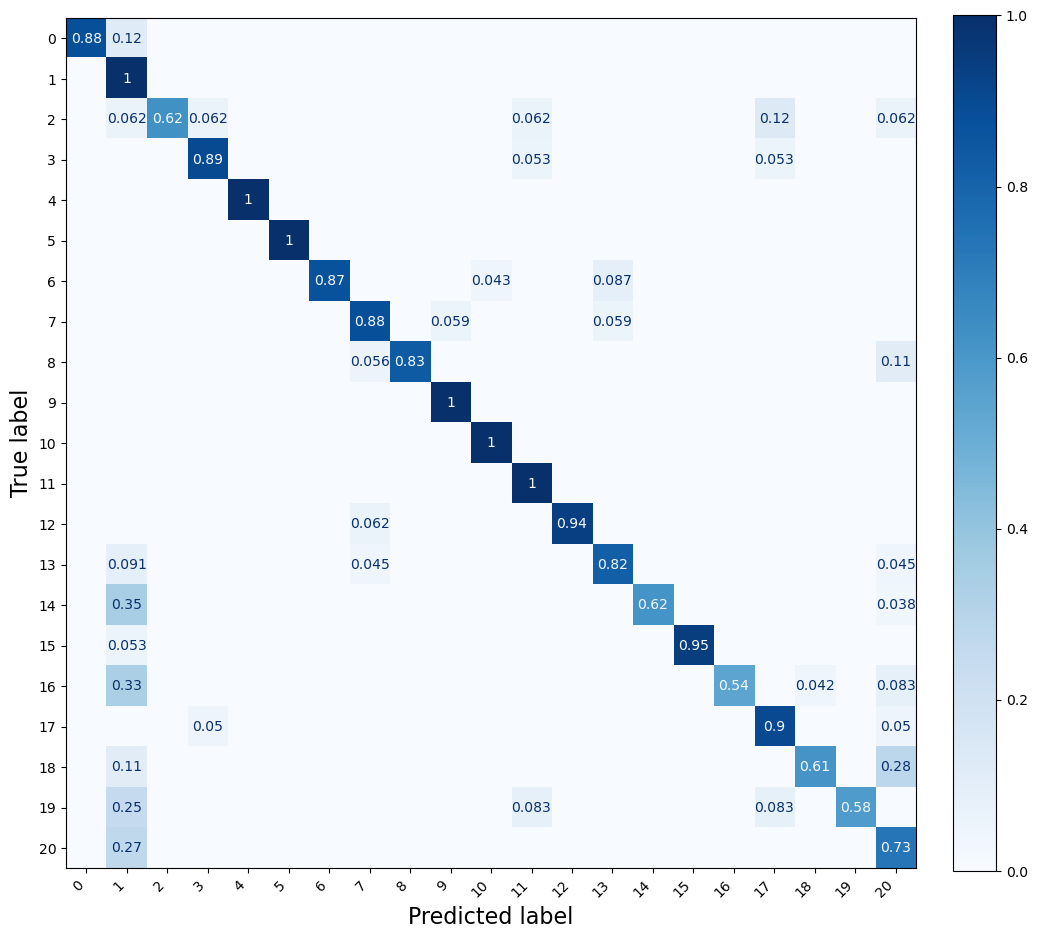}
    \caption{Accuracy=83.46\%}
    \label{Conf:sub8}
\end{subfigure}
\hfill
\begin{subfigure}{0.19\textwidth}
    \includegraphics[width=\linewidth]{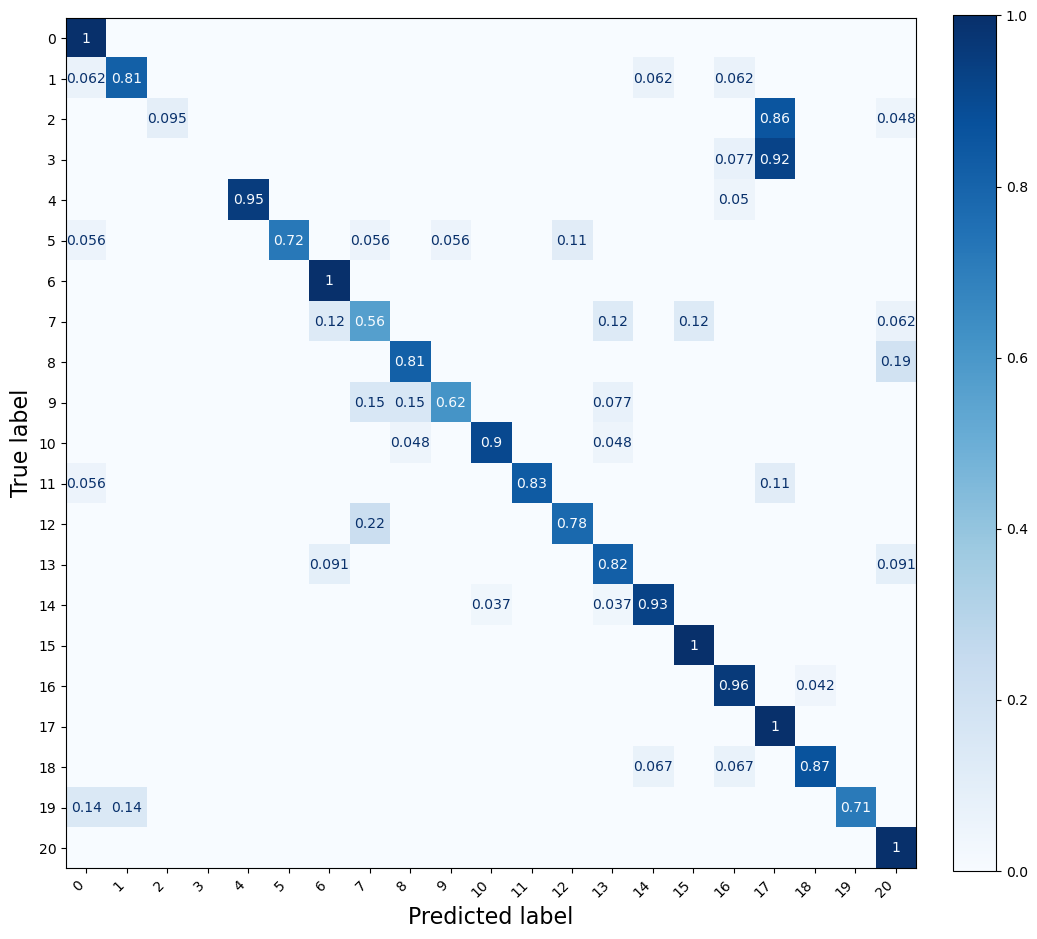}
    \caption{Accuracy=80.66\%}
    \label{Conf:sub9}
\end{subfigure}
\hfill
\begin{subfigure}{0.19\textwidth}
    \includegraphics[width=\linewidth]{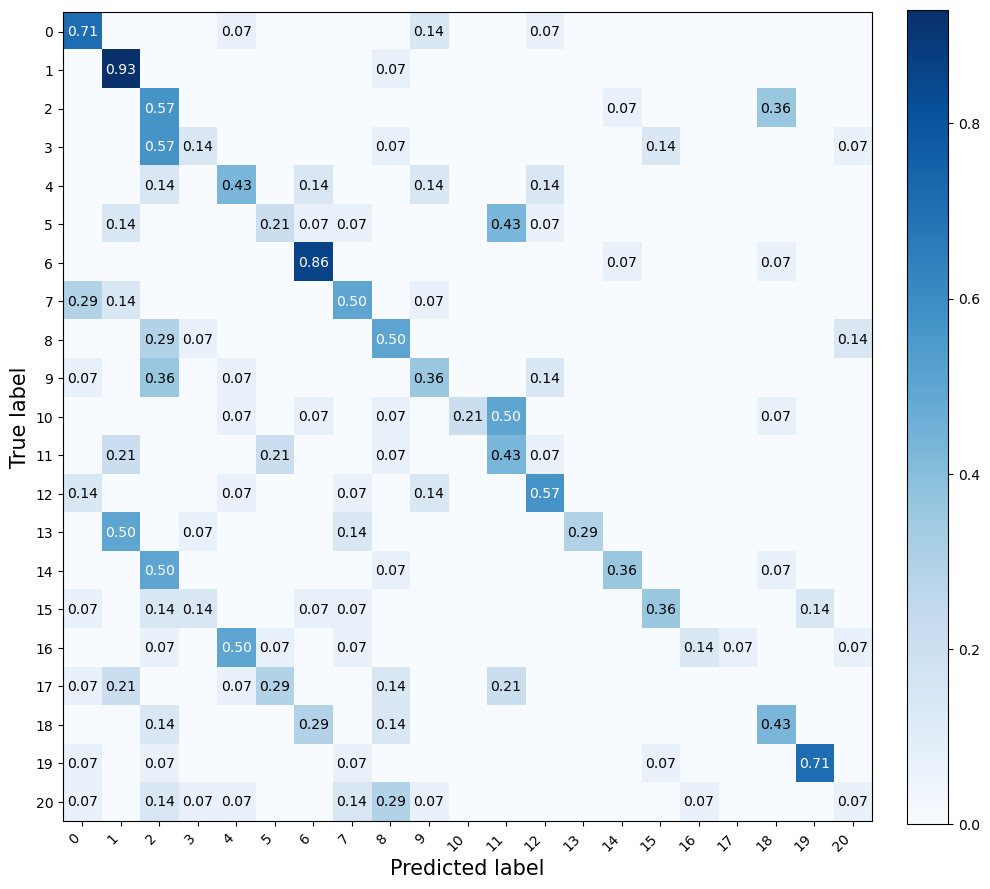}
    \caption{Accuracy=41.84\%}
    \label{Conf:sub10}
\end{subfigure}
\caption{Normalized confusion matrices obtained from models using phase, RSS, or both as input features: (a) RFC with SP; (b) RFC with SWP; (C) RFC with SPR; (d) SVM with SP; (e) SVM with SWP; (f) SVM with SPR; (g) Late Fusion; (h) Early Fusion; (i) EUIGR; (j) GRfid}
\label{ConfPhase}
\end{figure*}

\begin{figure*}[htbp]
\centering
\begin{subfigure}{0.19\textwidth}
    \includegraphics[width=\linewidth]{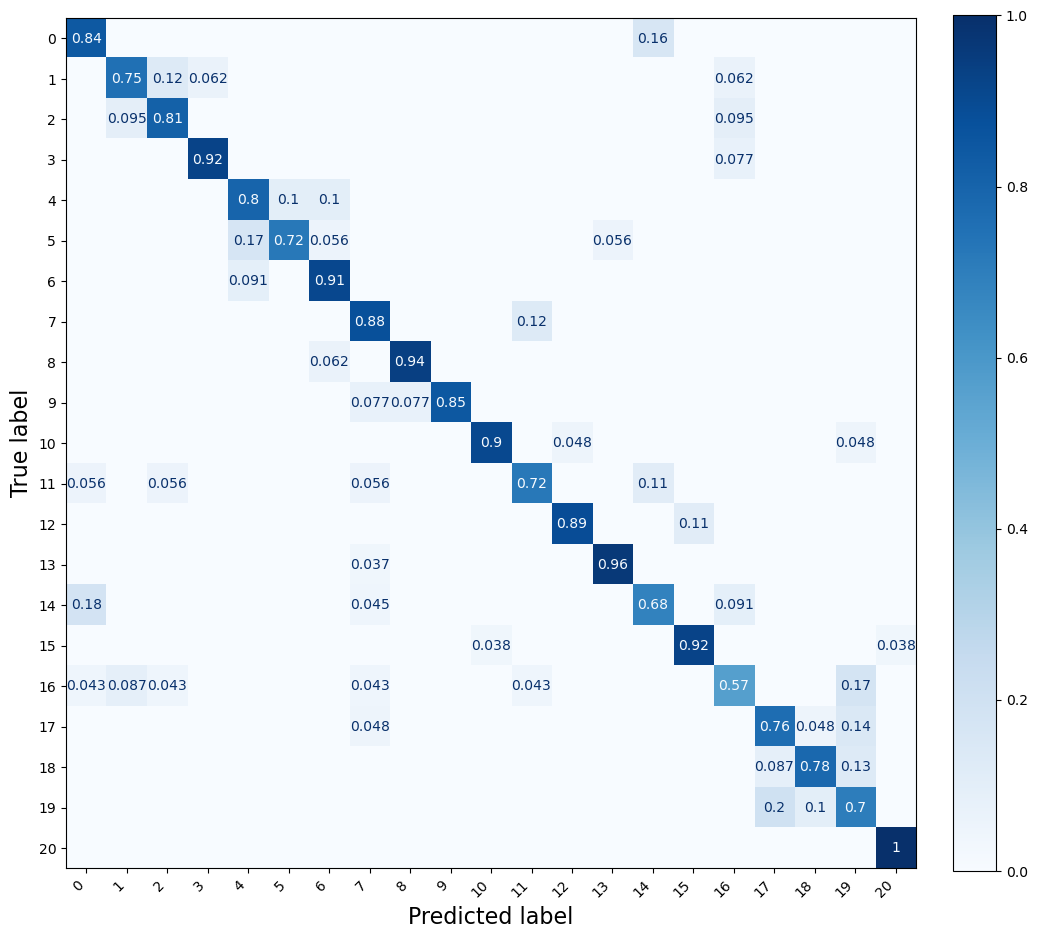}
    \caption{Accuracy=82.44\%}
    \label{ConfAoA:sub1}
\end{subfigure}
\hfill
\begin{subfigure}{0.19\textwidth}
    \includegraphics[width=\linewidth]{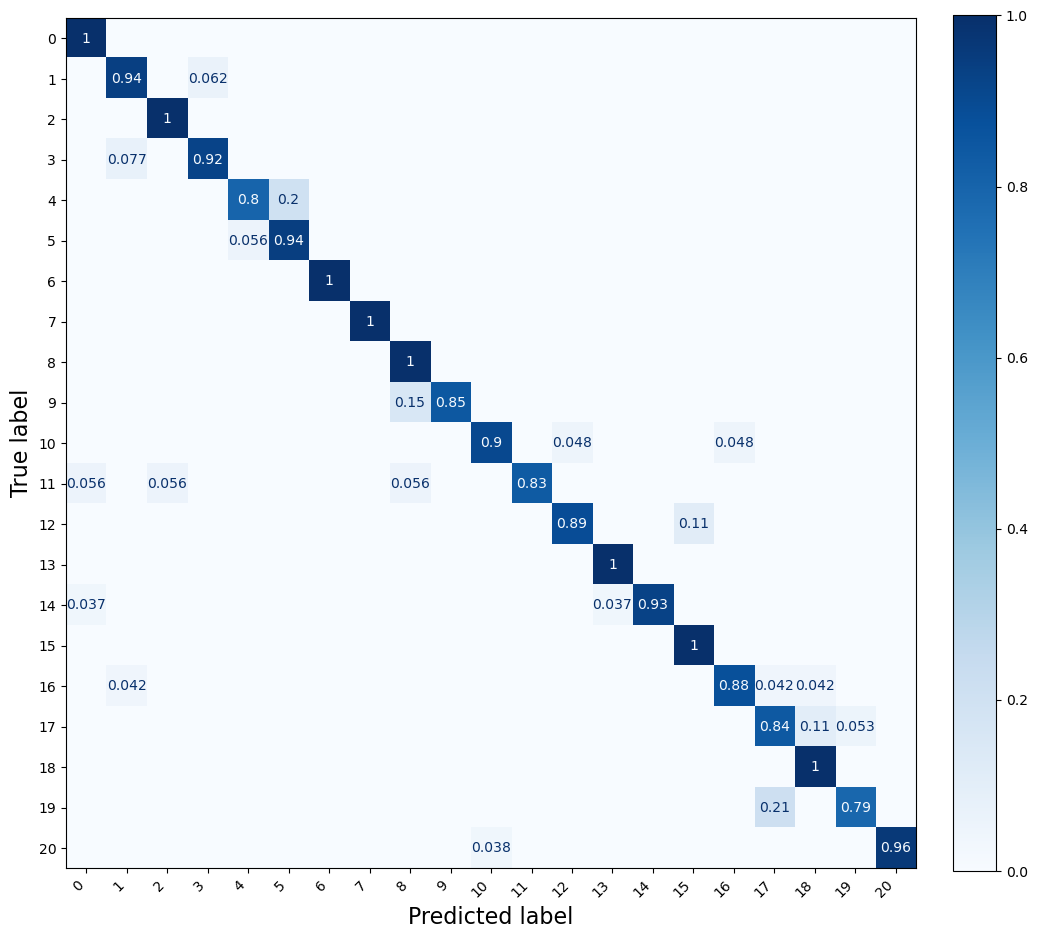}
    \caption{Accuracy=93.12\%}
    \label{ConfAoA:sub2}
\end{subfigure}
\hfill
\begin{subfigure}{0.19\textwidth}
    \includegraphics[width=\linewidth]{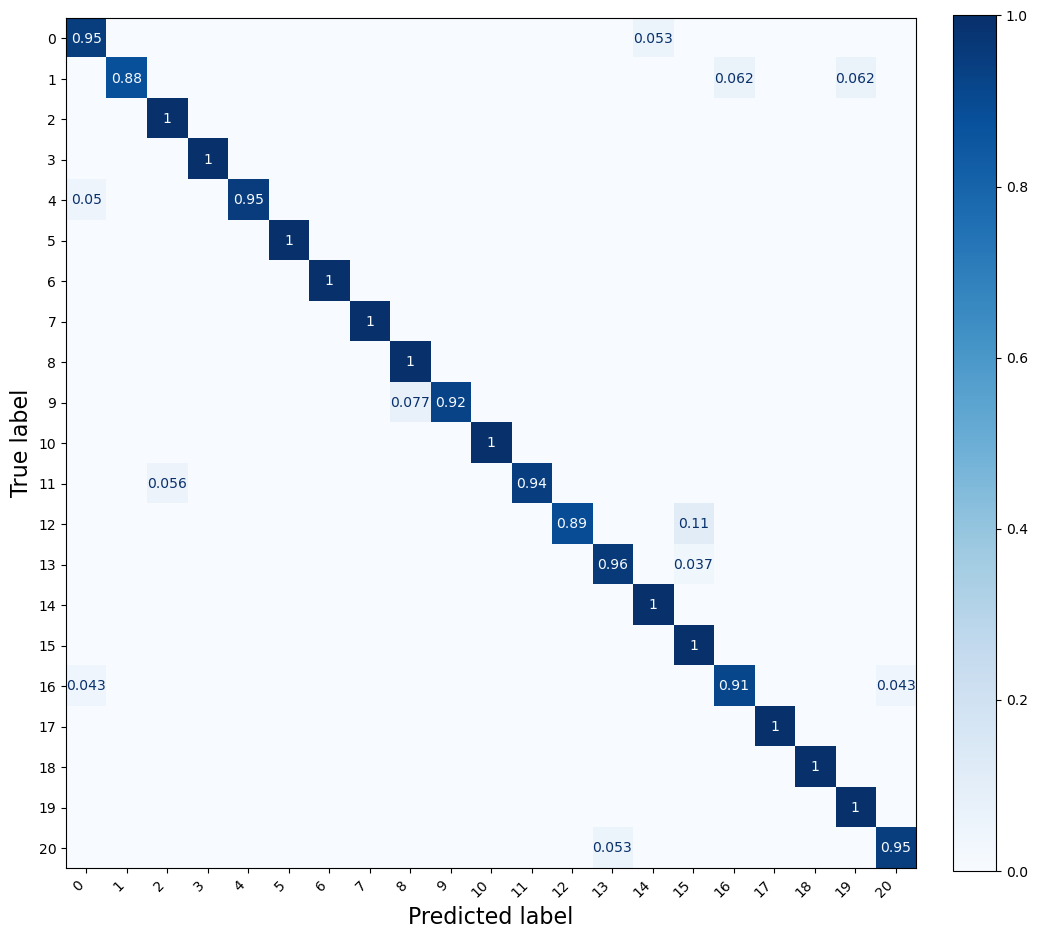}
    \caption{Accuracy=97.2\%}
    \label{ConfAoA:sub3}
\end{subfigure}
\hfill
\begin{subfigure}{0.19\textwidth}
    \includegraphics[width=\linewidth]{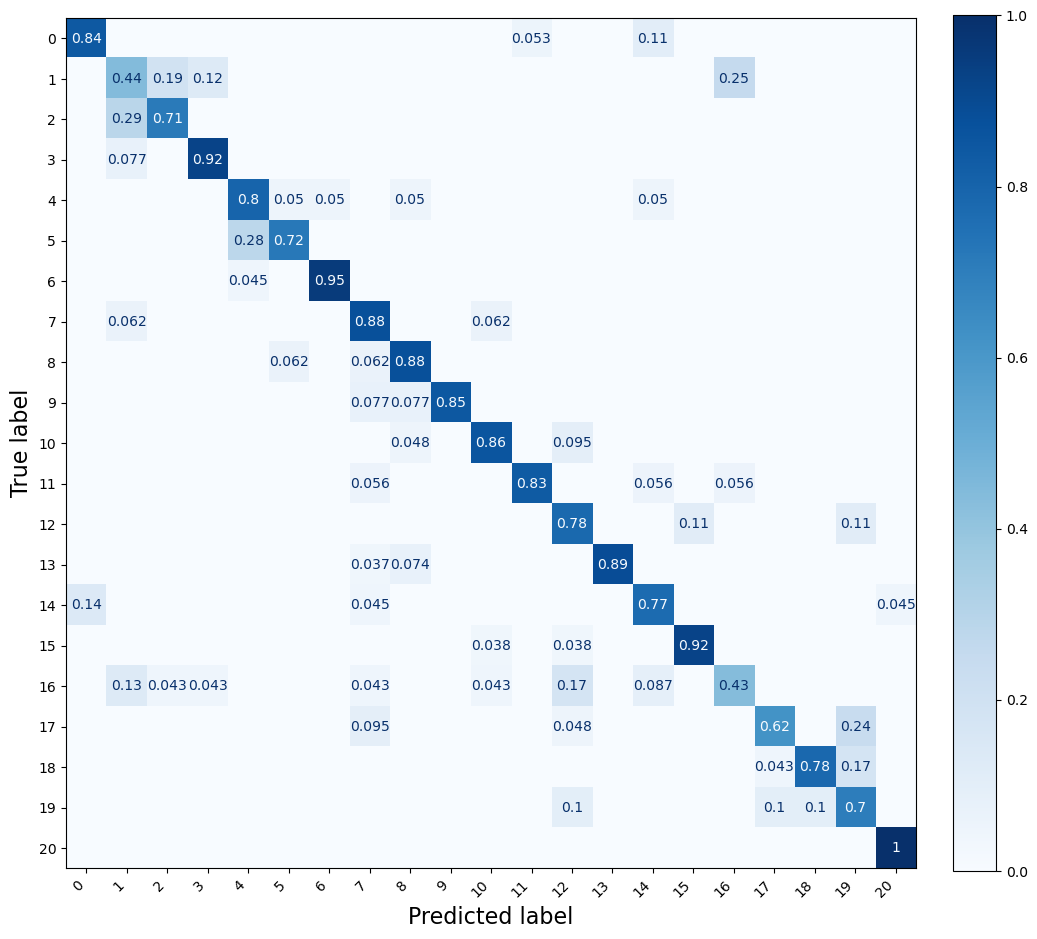}
    \caption{Accuracy=79.13\%}
    \label{ConfAoA:sub4}
\end{subfigure}
\hfill
\begin{subfigure}{0.19\textwidth}
    \includegraphics[width=\linewidth]{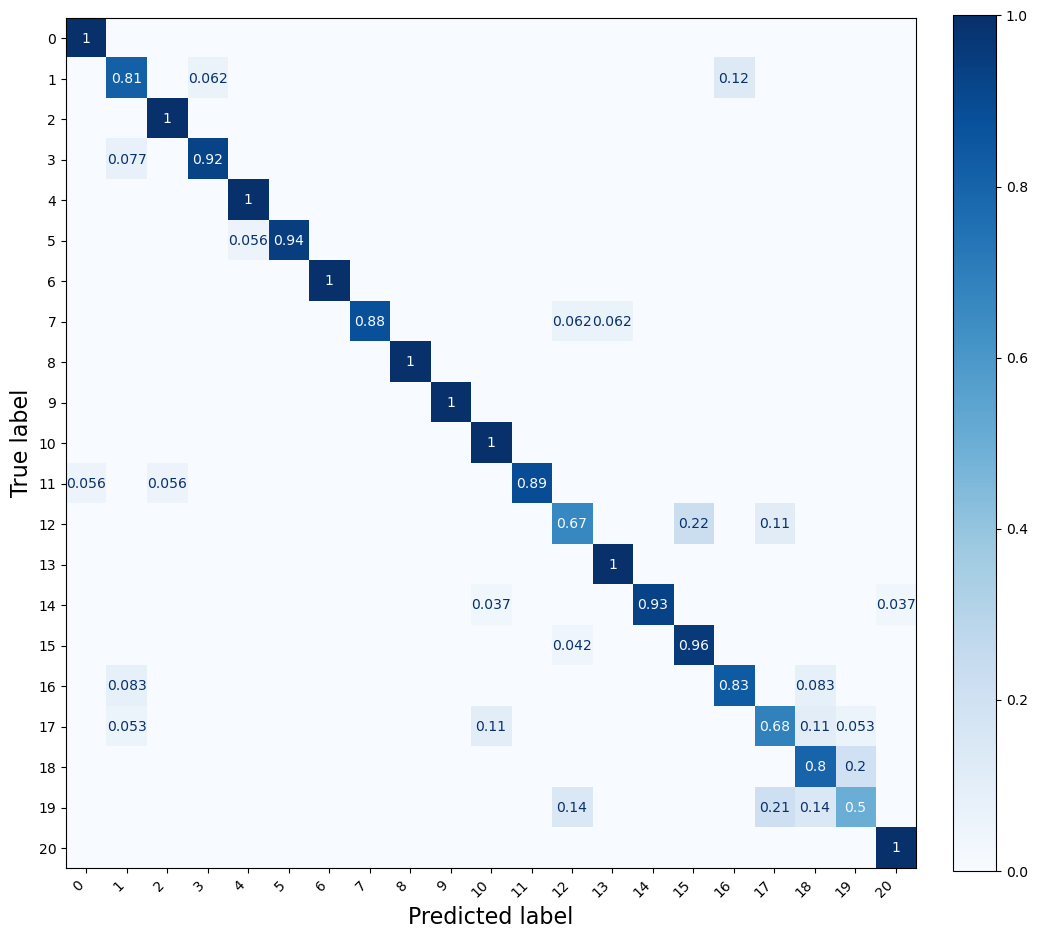}
    \caption{Accuracy=91.09\%}
    \label{ConfAoA:sub5}
\end{subfigure}

\begin{subfigure}{0.19\textwidth}
    \includegraphics[width=\linewidth]{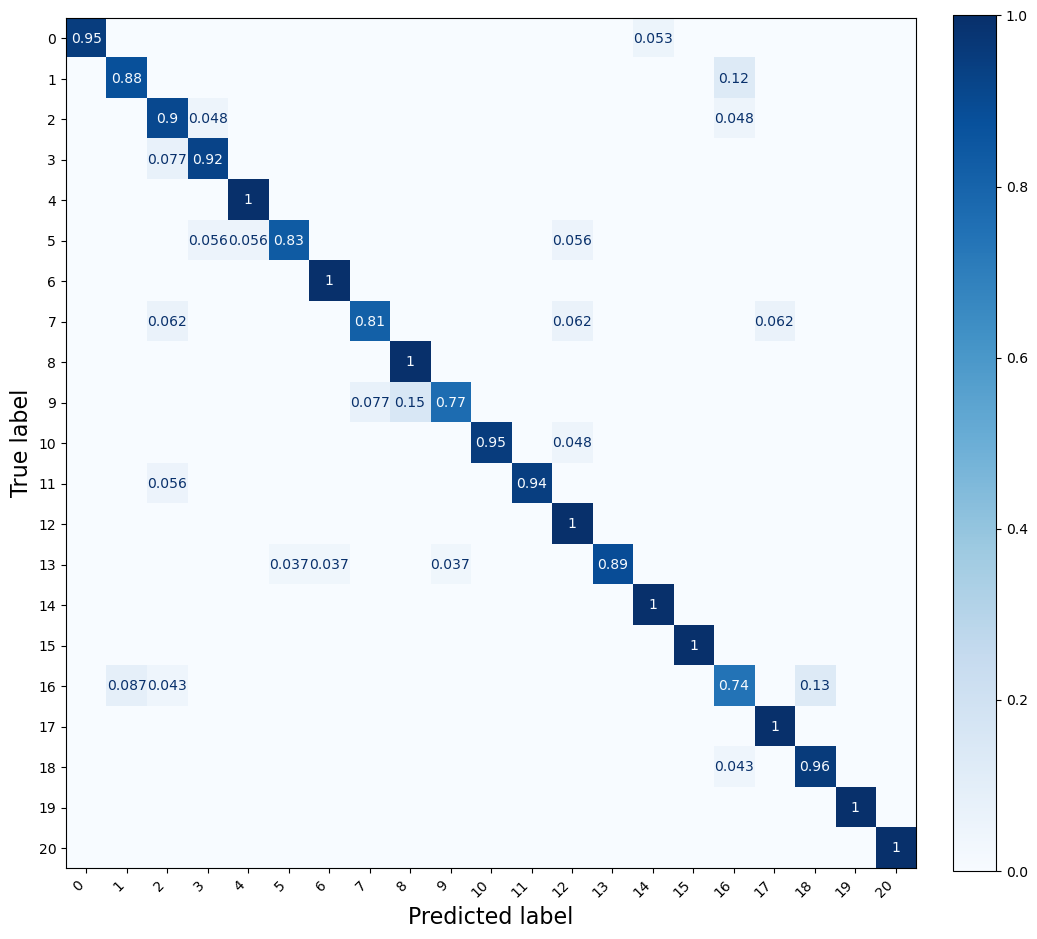}
    \caption{Accuracy=93.12\%}
    \label{ConfAoA:sub6}
\end{subfigure}
\hfill
\begin{subfigure}{0.19\textwidth}
    \includegraphics[width=\linewidth]{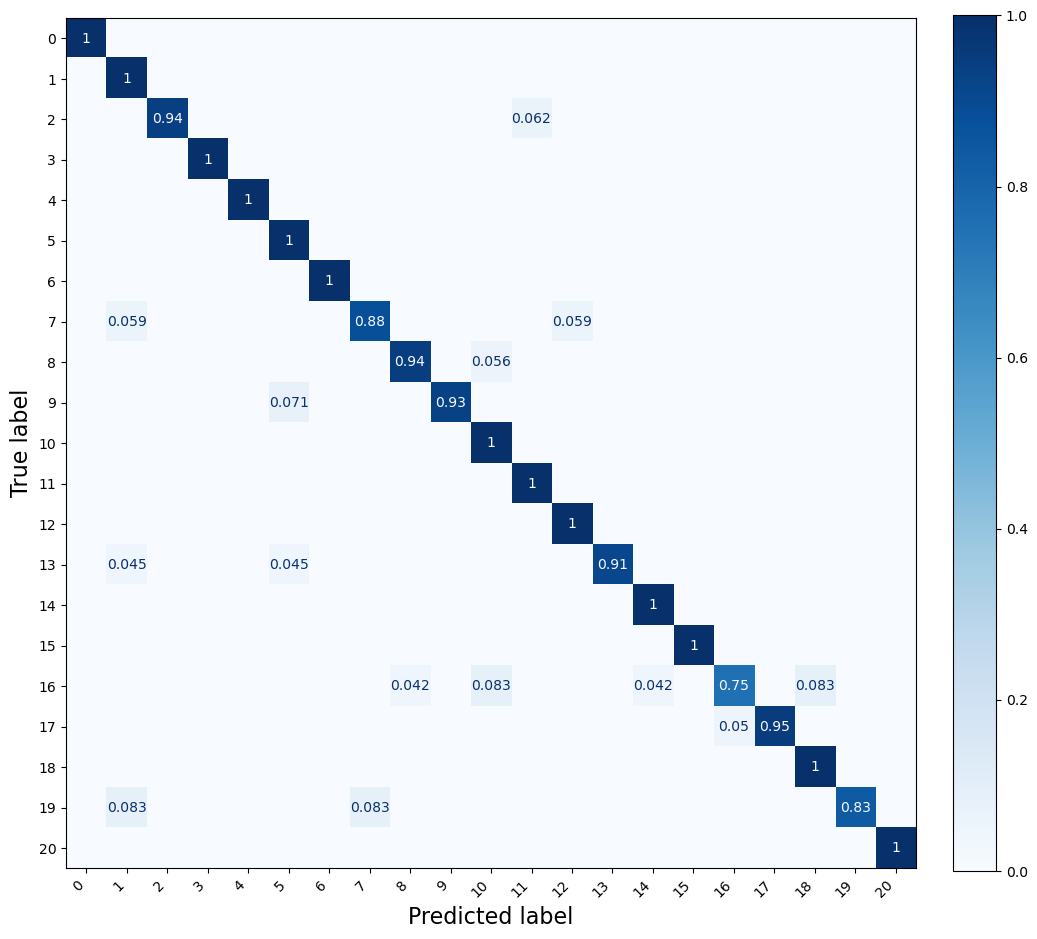}
    \caption{Accuracy=95.92\%}
    \label{ConfAoA:sub7}
\end{subfigure}
\hfill
\begin{subfigure}{0.19\textwidth}
    \includegraphics[width=\linewidth]{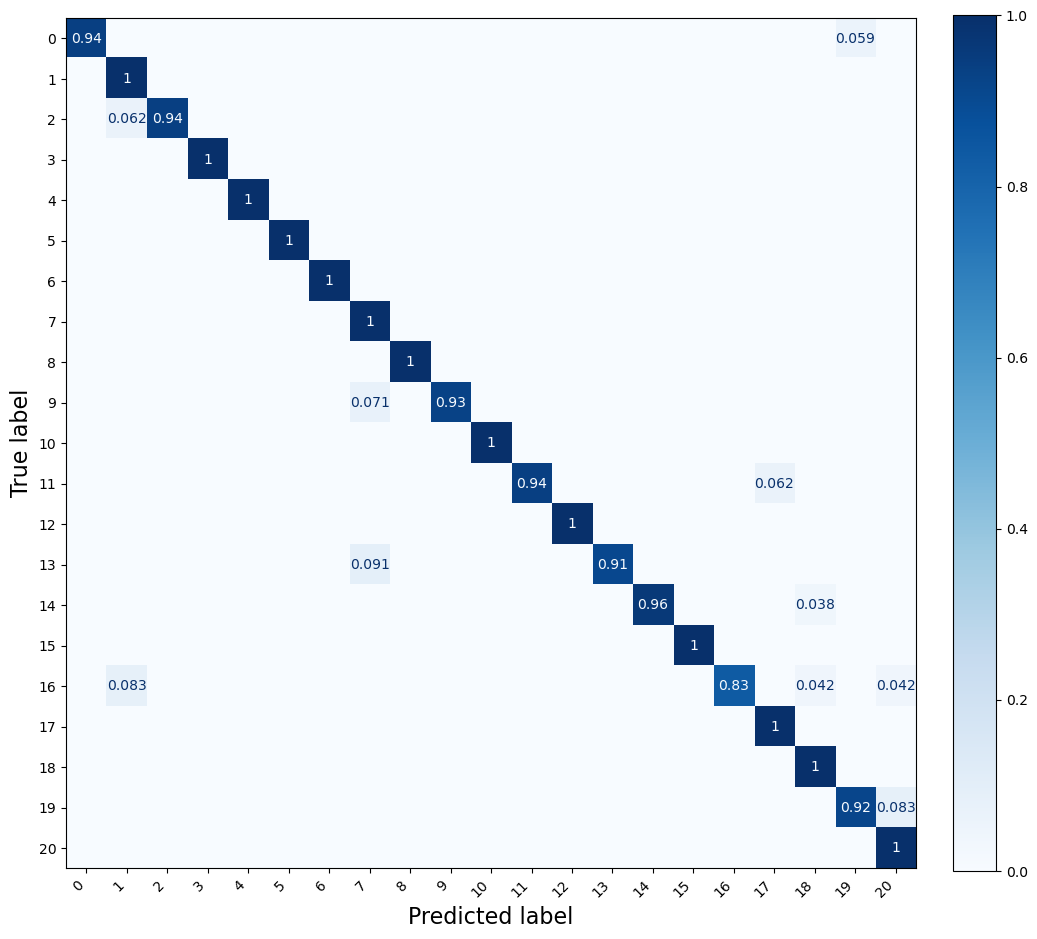}
    \caption{Accuracy=96.94\%}
    \label{ConfAoA:sub8}
\end{subfigure}
\hfill
\begin{subfigure}{0.19\textwidth}
    \includegraphics[width=\linewidth]{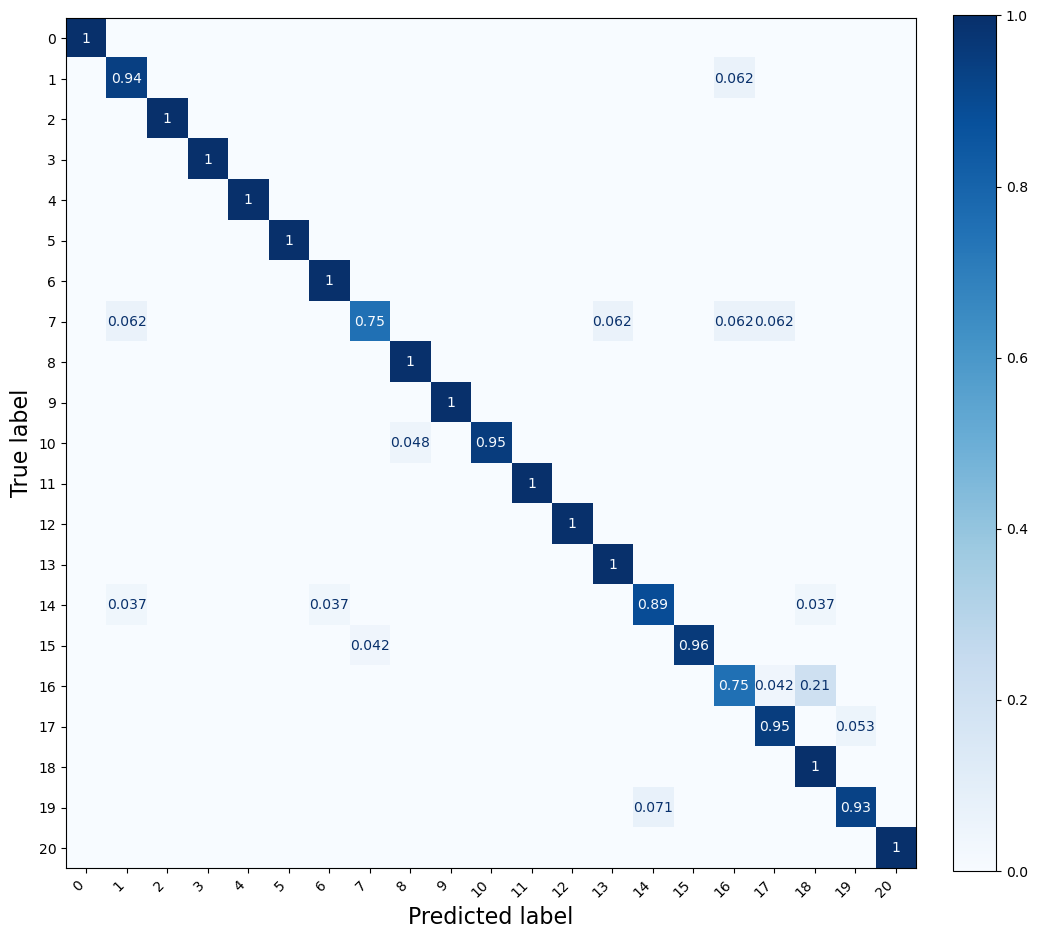}
    \caption{Accuracy=95.41\%}
    \label{ConfAoA:sub9}
\end{subfigure}
\hfill
\begin{subfigure}{0.19\textwidth}
    \includegraphics[width=\linewidth]{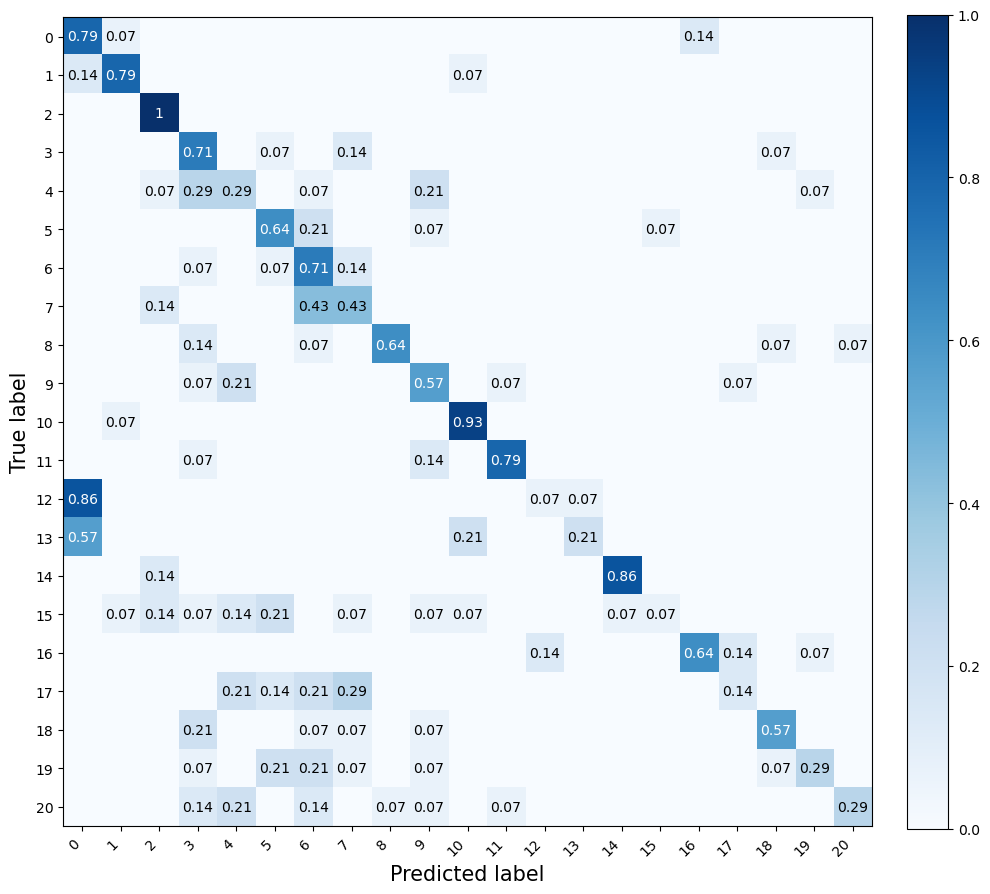}
    \caption{Accuracy=54.42\%}
    \label{ConfAoA:sub10}
\end{subfigure}
\caption{Normalized confusion matrices obtained from models using phase, RSS, AoA, or their combinations as input features.: (a) RFC with SA; (b) RFC with SWA; (C) RFC with SPRA; (d) SVM with SA; (e) SVM with SWA; (f) SVM with SPRA; (g) Late Fusion; (h) Early Fusion; (i) EUIGR; (j) GRfid}
\label{ConfAoA}
\end{figure*}

The normalized confusion matrices are shown in Fig~\ref{ConfPhase} for models that use phase or a combination of phase and RSS as input features, and in Fig~\ref{ConfAoA} for models that use AoA or its combination with phase and RSS. It can be observed that, while it is rare for any gesture to be classified with 100\% accuracy across all methods in Fig~\ref{ConfPhase}, it frequently occurs that some gestures achieve perfect classification in certain methods when AoA tracking is utilized. Overall, incorporating AoA into phase and RSS signals or using AoA tracking alone considerably boosts the performance of gesture recognition systems.

\begin{table}
\caption{Accuracy (Acc.), Precision (Pre.), Recall (Rec.), and F1-score (F1) of methods using phase and RSS inputs, excluding AoA.}   
\centering
\setlength\tabcolsep{5.1pt} 
\begin{tabular}{|c|c|c|c|c|}
\hline
\multicolumn{1}{|c|}{\diagbox[width=11em]{\textbf{Method}}{\textbf{Metric}}} &  \textbf{Acc.} & \textbf{Pre.} & \textbf{Rec.} & \textbf{F1} \\ \hline
RFC with \textbf{SP} & $82.18$ & $82.67$ & $82.18$ & $82.06$ \\ \hline
RFC with \textbf{SWP} & $84.22$ & $84.79$ & $84.22$ & $84.18$ \\ \hline
RFC with \textbf{SPR} & $85.49$ & $85.74$ & $85.48$ & $85.35$ \\ \hline
SVM with \textbf{SP} & $75.06$ & $76.3$ & $75.06$ & $75.15$ \\ \hline
SVM with \textbf{SWP} & $79.89$ & $80.62$ & $79.89$ & $79.69$ \\ \hline
SVM with \textbf{SPR} & $78.11$ & $79.36$ & $78.1$ & $78.09$ \\ \hline
Early Fusion  & $83.46$ & $89.73$ & $83.46$ & $84.45$ \\ \hline
Late Fusion  & $83.96$ & $85.31$ & $83.96$ & $83.92$ \\ \hline
EUIGR & $80.66$ & $83.11$ & $80.66$ & $78.65$ \\ \hline
GRfid & $41.84$ & $47.58$ & $41.83$ & $39.66$ \\ \hline
\end{tabular}
\label{tab1}
\end{table}


\begin{table}
\caption{Accuracy (Acc.), Precision (Pre.), Recall (Rec.), and F1-score (F1) of methods using phase, RSS, and AoA.}   
\centering
\setlength\tabcolsep{5.1pt} 
\begin{tabular}{|c|c|c|c|c|}
\hline
\multicolumn{1}{|c|}{\diagbox[width=11em]{\textbf{Method}}{\textbf{Metric}}} &  \textbf{Acc.} & \textbf{Pre.} & \textbf{Rec.} & \textbf{F1} \\ \hline
RFC with \textbf{SA} & $82.44$ & $83.56$ & $82.44$ & $82.67$ \\ \hline
RFC with \textbf{SWA} & $93.12$ & $93.56$ & $93.12$ & $93.11$ \\ \hline
RFC with \textbf{SPRA} & $97.2$ & $97.31$ & $97.2$ & $97.19$ \\ \hline
SVM with \textbf{SA} & $79.13$ & $81.38$ & $79.12$ & $79.52$ \\ \hline
SVM with \textbf{SWA} & $91.09$ & $91.23$ & $91.09$ & $91$ \\ \hline
SVM with \textbf{SPRA} & $93.12$ & $93.39$ & $93.12$ & $93.08$ \\ \hline
Early Fusion  & $96.94$ & $97.26$ & $96.94$ & $96.95$ \\ \hline
Late Fusion  & $95.92$ & $96.21$ & $95.92$ & $95.86$ \\ \hline
EUIGR & $95.41$ & $95.76$ & $95.41$ & $95.39$ \\ \hline
GRfid & $54.42$ & $59.18$ & $54.42$ & $52.15$ \\ \hline
\end{tabular}
\label{tab2}
\end{table}

\section{Conclusion}
In this work, we addressed the challenge of recognizing a large variety of hand gestures using passive reflective tags by introducing AoA tracking as a complementary feature to conventional RSS and phase-based methods. We showed that RSS and phase signals often lack the discriminative power required for complex gesture differentiation due to signal similarity. By using AoA, which varies with hand movement, we were able to capture spatial information that enhanced the separability of gestures.
Through the use of the MUSIC algorithm and Smart Antenna Switching (SAS), we demonstrated AoA estimation in static tag configurations. We then developed a Kalman smoothing-based AoA tracking algorithm for gesture execution. Our experiments confirmed that incorporating AoA features into gesture recognition pipelines leads to substantial improvements in classification accuracy, with gains of up to 15\% in benchmark tests.
These findings highlight the value of AoA as a robust and distinguishing feature for RF-based gesture recognition.


%

\appendices


\ifCLASSOPTIONcompsoc
\else


\ifCLASSOPTIONcaptionsoff
  \newpage
\fi

\balance 
\bibliographystyle{ieeetr}
\bibliography{ref}

\end{document}